\documentclass[aps,pra,nobalancelastpage,superscriptaddress,twocolumn,10pt]{revtex4-2}
\usepackage{mathtools}

\usepackage{color,ifthen,amsthm,amsmath,amsxtra,amsfonts,dsfont,graphicx,bm,tikz,scalerel,wasysym,bbm,amsthm,amssymb,braket,physics,empheq,CircuitTikz, comment,thmtools,thm-restate}
\usepackage[colorlinks=true,linkcolor=blue, citecolor=blue, urlcolor=blue, bookmarks]{hyperref}

\def\Id{{\openone}}
\newcommand{\be}{\begin{equation}}
\newcommand{\ee}{\end{equation}}
\newcommand{\bea}{\begin{eqnarray}}
\newcommand{\eea}{\end{eqnarray}}
\newcommand{\bse}{\begin{subequations}}
\newcommand{\ese}{\end{subequations}}
\newcommand{\prlsection}[1]{{\em {#1}.---}}

\newcommand{\otimesR}{\mkern-4mu\otimes_{\mkern-2mu \rm r}\mkern-4mu}
\newcommand{\ketR}[1]{\ket{#1}_{\mkern-4mu \rm r}}

\newcommand{\ketbraR}[2]{\ketR{#1}\mkern-7mu\bra{#2}}

\newcommand{\Rbraket}[2]{\prescript{}{\rm r\mkern-5mu}{\braket*{#1}{#2}}}

\theoremstyle{plain}

\theoremstyle{plain}

\theoremstyle{plain}

\theoremstyle{plain}

\begin{document}

\title{Random Permutation Circuits Beyond Qubits are Quantum Chaotic}

\author{Bruno Bertini}
\author{Katja Klobas}
\affiliation{School of Physics and Astronomy, University of Birmingham, Birmingham, B15 2TT, UK}
\author{Pavel Kos}
\affiliation{Max-Planck-Institut f\"ur Quantenoptik, Hans-Kopfermann-Str.~1, 85748 Garching, Germany}

\author{Daniel Malz}
\affiliation{Department of Mathematical Sciences, University of Copenhagen, 2100 Copenhagen, Denmark}

\begin{abstract}
Random permutation circuits were recently introduced as minimal models for local many-body dynamics that can be interpreted both as classical and quantum. Standard dynamical complexity indicators such as damage spreading and out-of-time-order correlators (OTOCs), show that these systems exhibit sensitivity to initial conditions in the classical setting and operator scrambling in the quantum setting. Here, we address their quantum chaoticity --- a stricter property --- by studying the time evolution of local operator entanglement (LOE). We show that the behaviour of LOE in random permutation circuits depends on the dimension of the local configuration space $q$. When $q=2$, i.e.\ the circuits act on qubits, random permutations are Clifford and the LOE of any local operator is bounded by a constant, indicating that they are not truly chaotic. On the other hand, when the dimension of the local configuration space exceeds two, the LOE grows linearly in time. We prove this in the limit of large $q$ and present numerical evidence that a three-dimensional local configuration space is sufficient for a linear growth of LOE. Our findings highlight that quantum chaos can be produced by essentially classical dynamics. Moreover, we show that LOE can be defined also in the classical realm and put it forward as a universal indicator chaos, both quantum and classical. 
\end{abstract}
\maketitle

\footnotetext[20]{See the Supplemental Material for: (i) A proof of Eq.~\eqref{eq:PuritySimp}; (ii) Further details on the permutation-averaged multi-replica gates; (iii) The discussion of dynamics of permutation gates with added phases; (iv) The perturbative large local-Hilbert-space expansion; (v) Additional details on the numerical methods to produce the data in the main text. }

\prlsection{Introduction}Quantum many-body systems in discrete spacetime, also known as quantum circuits, have emerged in recent years as a fruitful arena to study quantum non-equilibrium dynamics~\cite{potter2022entanglement,fisher2022random,bertini2025exactly}. In particular, random unitary circuits (RUC), where the evolution is generated by uncorrelated random unitary matrices coupling nearest neighbours, have proved to be especially instructive~\cite{nahum2017quantum,chan2018solution,nahum2018operator,vonKeyserlingk2018operator,zhou2019emergent,bertini2020scrambling,fisher2022random}. In essence, this is because (i) RUC have minimal structure (only locality is retained in the most basic setting) and thus can be used to study the universal aspects of quantum information spreading, and (ii) certain key properties of these systems can be derived analytically, employing the so called Weingarten calculus to compute averages over the unitary group~\cite{collins2022weingarten}. The same basic idea was recently adopted also to study classical many-body dynamics and understand how it differs from its quantum counterpart~\cite{iaconis2019anomalous, iaconis2021quantum, han2023entanglement, feng2025dynamics, bertini2025quantum, szaszschagrin2025entanglement, mcdonough2025bridging}. In this case one considers a subset of random unitary circuits, dubbed random permutation circuits (RPC)~\cite{bertini2025quantum}, that in a special basis act as permutations of the basis elements and hence `classically' (a similar class is that of random automaton circuits~\cite{iaconis2019anomalous}, which, besides permuting the special basis elements, also multiplies them by phases). The key feature of these systems is that they can be thought of both as quantum circuits and as classical cellular automata or Boolean networks~\cite{kadanoff2002boolean}. In this way, they allow for a quantitative comparison between quantum and classical dynamics~\cite{Pizzi2022,Pizzi2024}, avoiding the confusion arising when comparing completely different setups (e.g.\ quantum spin chains and few-particle classical systems with a continuous phase space). 


A natural application of these systems is to the study of chaos, whose quantum and classical versions are currently treated on very different footings. In essence, the problem is that for generic quantum systems there is no analogue to the familiar concept of ``sensitivity to initial conditions'', which defines classical chaos. An analogue exists in the semiclassical regime (or, more generally, in the presence of a large parameter identifying a small effective Planck constant $h_{\rm eff}\ll 1$) where it is quantified by out-of-time-ordered correlators (OTOCs)~\cite{shenker2014black, shenker2014multiple, roberts2015localized, polchinski2016spectrum, maldacena2016bound}, but not when $h_{\rm eff}\sim 1$. In the latter case OTOCs only measure ``operator scrambling'', not chaos, meaning that they are unable to distinguish regular and non-regular dynamics~\cite{dowling2023scrambling}. The current belief is that, besides the well established spectral statistics~\cite{casati1980on, bohigas1984characterisation}, another diagnostic of quantum chaos is provided  by \emph{local operator entanglement} (LOE)~\cite{prosen2007operator,prosen2007is,Pizorn_2009,jonay2018coarse,alba2019operator,alba2021diffusion,bertini2020operatorI,bertini2020operatorII,dowling2022operational,Wellnitz_2022,Kudler_Flam_2020,Rath_2023,alba2024operatorspaceentanglementose}: both these concepts, however, have no obvious classical analogue. One can hope that, given their dual `quantum-classical' nature, RPC can help resolving this dissatisfactory situation and establish a unified description of chaos encompassing both classical and quantum systems. So far, however, only their sensitivity to initial conditions in the classical setting (via the so called ``damage spreading''~\cite{bernard2011damage,das2018lightcone,liu2021butterfly}) and their operator scrambling (via OTOCs) in the quantum setting have been rigorously characterised~\cite{bertini2025quantum}, whereas the current results concerning their quantum chaoticity are not fully conclusive, as they rely on indirect arguments of small-system numerics~\cite{iaconis2019anomalous, iaconis2021quantum, mcdonough2025bridging}.

Here we report key developments on this matter. First, we characterise the quantum chaotic properties of RPC. Specifically, by performing a rigorous perturbative expansion in the local Hilbert space dimension $q$, we prove that RPC with nearest-neighbour interactions generate a linearly growing LOE of diagonal operators in the limit of large $q$.
Next, we extend this result to general operators via a mapping to an effective statistical mechanical model, and present numerical evidence suggesting that the linear growth persists for all $q>2$. On the other hand, we show that for $q=2$ LOE is bounded as the circuits belong to the Clifford group~\cite{nielsen2000quantum}. This difference is not captured by damage spreading or OTOCs and showcases the difference between these diagnostics. Finally, noting that LOE of diagonal operators is well defined in the classical realm and is more stringent than damage spreading, we propose it as a universal diagnostic tool for chaos applicable not only for quantum but also for classical systems.

\prlsection{Setting}We consider $2L$ qudits with $q$ internal states each arranged along a one-dimensional lattice with the sites labelled by the elements of $\mathbb Z_{2L}/2=\{0,1/2, \ldots, L-1/2\}$. The time evolution proceeds in discrete steps and is implemented by the unitary operator $\mathbb{U}(\tau)=\mathbb{U}_{\rm e}(\tau)\mathbb{U}_{\rm o}(\tau)$, where
\begin{equation}
\label{eq:brickwork}
  \mathbb{U}_{\rm o}(\tau)= \bigotimes_{x=1}^{L} U(x,\tau-\tfrac{1}{2}) , \quad \mathbb{U}_{\rm e}(\tau)=\Pi^{\dagger}  \mathbb{U}_{\rm o}(\tau+\tfrac{1}{2}) \Pi,
\end{equation}
$\Pi$ implements a periodic one-site shift, and $U(x,\tau)$ are local two-site time-evolution operators, or \emph{gates}, acting at time $\tau$ on the qudits at sites $x-1/2$ and $x$. This defines a `brickwock' quantum circuit implementing local dynamics.

Our focus is on the case where $U(x,\tau)$ are chosen so that there exists a fixed special \emph{local} basis (computational basis) $\{\ket{n}\}_{n=0}^{q-1}$, in which the gates act as permutations of the $q^2$ basis elements. Introducing the shorthand notation for a two-qudit state $\ket{n}=\ket{n_1 n_2}$, where $n=q n_1+n_2$ and $n_{1,2}\in\mathbb{Z}_q$, $U(x,\tau)$ can be expressed as 
\begin{equation}
\label{eq:permutationproperty}
U(x,\tau) \ket{n} 
  = 
  \ket{\pi_{x,\tau}(n)}, \qquad \pi_{x,\tau}\in S_{q^2}\,,
\end{equation}
where $S_{q^2}$ is the group of permutations of $q^2$ elements. The resulting circuit is a permutation circuit (a special case of an \emph{automaton circuit}~\cite{iaconis2019anomalous}) and does not generate entanglement in the computational basis. Thus, the dynamics of the system can be thought of as classical: we can associate to the quantum circuit an equivalent classical reversible cellular automaton (or boolean circuit) evolving a classical configuration $(s_0, s_{1/2}, \ldots, s_{L-1/2})$ in the same way as the quantum circuit evolves the basis state $\ket*{s_0 s_{1/2} \ldots s_{L-1/2}}$.

To compute local operator entanglement (LOE), we consider the local operator $\mathcal{O}_x$, acting non-trivially as a $q\times q$ operator $\mathcal{O}$ at site $x$, and as the identity otherwise. We take $\mathcal{O}$ to be traceless, ${\tr[\mathcal{O}]=0}$, and compute its evolution in the Heisenberg picture $\mathcal{O}_x(t) =  \mathcal U(t)^{-1} \mathcal{O}_x \mathcal U(t)$,  where
\be
\mathcal U(t) \coloneqq \mathbb{U}(t) \cdots \mathbb{U}(1). 
\ee
Next, following Refs.~\cite{prosen2007operator,prosen2007is}, we compute the ``entanglement'' of $\mathcal{O}_x(t)$ by viewing it as a quantum state in a doubled Hilbert space, $\ket{\mathcal{O}_x(t)}$, using the mapping
\begin{equation}
  \ketbra{s_1 s_2 \ldots s_{2L}}{r_1 r_2\ldots r_{2L}} \mapsto
  \ket{s_1 r_1 s_2 r_2 \ldots s_{2L} r_{2L}},
\end{equation}
which establishes a correspondence between operators acting on qudits with local dimension $q$ and states of qudits with local dimension $q^2$. Then, we estimate its entanglement over a spatial bipartition, $\mathbb Z_{2L}/2= \bar{A}\cup A$. To this end, we compute the second R\'enyi entropy of $\ket{\mathcal{O}_{L/2}(t)}$, reduced to the subsystem $A$. Choosing the normalisation $\tr\smash{[\mathcal{O}\mathcal{O}^{\dagger}]}=q$, we obtain 
\begin{equation}
\label{eq:LOE}
  S_{\mathcal{O},2} (t)= - \log \tr[\rho_{\mathcal{O},A}^{2}(t)] \eqqcolon - \log \mathcal{P}_{\mathcal{O}}(t),
\end{equation}
where we set $\rho_{\mathcal{O},A}(t)\coloneqq q^{-2L}\tr_{\bar{A}}[\ketbra{\mathcal{O}_{L/2}(t)}{\mathcal{O}_{L/2}(t)}]$ and introduced its purity $\mathcal P_{\mathcal{O}}(t)$. We focus on the case of a contiguous bipartition, with $\bar{A}$ containing the first $L+2\ell$ sites, and $A$ the remaining $L-2\ell$, with $\ell\ge 0$.  This quantity, sometimes called LOE R\'enyi entropy, has been proposed as a means to assess the ergodicity properties of systems in which the usual definition of quantum chaos through the spectral statistics (see, e.g., Refs.~\cite{Haake2019Quantum,kos2018many,chan2018solution}) cannot be applied due to the lack of a fixed propagator. In essence, its linear growth in the thermodynamic limit is associated with chaotic dynamics.

The purity in Eq.~\eqref{eq:LOE} can be represented as a matrix element in a replicated space via a standard \emph{folding} mapping (see e.g.\ Ref.~\cite{bertini2025exactly}). More specifically, we introduce the following states in the $8$-replica space of a single qudit 
\begin{equation}\label{eq:defStates8}
  \mkern-17mu
  \begin{aligned}
    \ket*{\circleSA^{(8)}\mkern-2mu}\mkern-6mu&=\mkern-8mu
    \ket*{\circleSA^{(4)}\mkern-2mu}\mkern-2mu\otimesR\mkern-4mu
    \ket*{\circleSA^{(4)}\mkern-2mu}\!,\mkern4mu
    \ket*{\blackcircleSA{\mathcal{O}}^{(8)}\mkern-2mu}\mkern-6mu 
    =
    \mkern-6mu\left(\mkern-2mu
    \mathcal{O} \otimesR \Id \otimesR \mathcal{O}^{\dagger} \otimesR \Id
    \mkern-2mu\right)^{\otimesR\mkern4mu 2}\mkern-6mu\ket*{\circleSA^{(8)}\mkern-2mu}\!,\\
    \ket*{\triangleSA^{(8)}\mkern-2mu}
    \mkern-6mu&=\mkern-6mu\frac{1}{q^2}
    \smashoperator{\sum_{a,b,c,d=0}^{q-1}}\ketR{abcddcba}\!,\qquad
    \ket*{\squareSA^{(8)}\mkern-2mu}\mkern-6mu=\mkern-8mu
    \ket*{\squareSA^{(4)}\mkern-2mu}\mkern-2mu\otimesR\mkern-4mu
    \ket*{\squareSA^{(4)}\mkern-2mu}\!,
  \end{aligned}
  \mkern-20mu
\end{equation}
where $\otimesR$ denotes the tensor product among different replicas, we use the shorthand notation $\ketR{a_1a_2 \ldots a_8}=\ket{a_1}\otimesR\cdots \otimesR \ket{a_8}$, and
\begin{equation}
\ket*{\squareSA^{(4)}}=\frac{1}{q}\smashoperator{\sum_{a,b=0}^{q-1}}\ketR{abba}\!,\quad
\ket*{\circleSA^{(4)}}=\frac{1}{q}\smashoperator{\sum_{a,b=0}^{q-1}}\ketR{aabb}\!
\end{equation}
are states in the $4$-replica space. In terms of these vectors we can write the purity as 
\begin{equation}\label{eq:PurityGeneral}
\mathcal{P}_{\mathcal{O}}(t)=q^{4L}
  \mel*{ \squareSA^{(8)}_{L+2\ell} \triangleSA^{(8)}_{L-2\ell}}{\mathcal{U}_8(t) }{\circleSA^{(8)}_{L}
  \mkern-4mu\blackcircleSA{\mathcal{O}}^{(8)}\mkern-4mu
  \circleSA^{(8)}_{L-1}},
\end{equation}
where we introduced the replicated time evolution operator
$\mathcal U_8(t) =(\mathcal U(t)\otimesR\mathcal U(t)^*\mkern-3mu)^{\,\otimesR\mkern6mu 4}
=\mathcal U(t)^{\,\otimesR\mkern6mu 8}$, 
used $\squareSA^{(8)}_{x}$ to denote a sequence of $x$ symbols $\squareSA^{(8)}$ (analogously for the others), and adopted the convention that sequences of symbols now denote tensor products in space, e.g.\ $\ket{\smash{\squareSA^{(8)}_{2} \triangleSA^{(8)}_{1}}}=\ket{\smash{\squareSA^{(8)} \squareSA^{(8)} \triangleSA^{(8)}}} = \ket{\smash{\squareSA^{(8)}}}\mkern-4mu\otimes\mkern-4mu\ket{\smash{\squareSA^{(8)}}}\mkern-4mu\otimes\mkern-4mu\ket{\smash{\triangleSA^{(8)}}}$.

While Eq.~\eqref{eq:PurityGeneral} works for any unitary $\mathcal U$, the expression can be simplified for automaton circuits when $\mathcal{O}$ is diagonal in the computational basis~\cite{iaconis2019anomalous}, in which case it reduces to an expectation value in the $4$-replica space
\begin{equation}
\label{eq:PuritySimp}
\mkern-4mu \mathcal{P}_{\mathcal{O}}(t) \mkern-4mu =  \mkern-2mu q^{2L} \mkern-2mu
  \mkern-4mu \mel*{ \circleSA^{(4)}_{{L}+2\ell} \squareSA^{(4)}_{{L}-2\ell}}{
    \mathcal U_4(t)}{\flatSA_{L}^{(4)} 
  \mkern-4mu\blackflatSA{(4)}_{\mathcal{O}}\mkern-4mu
  \flatSA^{(4)}_{L-1}}\mkern-4mu,
\end{equation}
where we introduced 
\begin{equation}\label{eq:defStates4}
\mkern-12mu\ket*{\flatSA^{(4)}} =\frac{1}{q^2} \smashoperator{\sum_{a,b,c,d=0}^{q-1}}
\ketR{abcd}\!, \,\,
 \ket*{\blackflatSA{(4)}_{\mathcal{O}}}
  = \left(\!\mathcal{O}\otimesR \mathcal{O}^{\dagger}\!\right)^{\otimesR\mkern6mu 2}
 \mkern-8mu\ket*{\flatSA^{(4)}}\!. 
 \mkern-4mu
\end{equation}
The first represents a flat sum of states in the 4-replica space and the second is orthogonal to it because of the traceless property of $\mathcal O$. A proof of Eq.~\eqref{eq:PuritySimp} is contained in the Supplemental Material (SM)~\cite{Note20}.

Equation \eqref{eq:PuritySimp} is well defined also in the classical setting and corresponds to the following experiment. Take four copies of the system and initialise the first two in random initial states. For $x\leq L+\ell$, initialise the third copy to agree with the first and the fourth with the second; and analogously for $x>L+\ell$, but with the roles of third and fourth copy exchanged. Then evolve all copies under the same evolution $\mathcal U$. In the end, record the state of all four copies on site $L/2$, and compute the product of values of $\mathcal{O}$ in each of the four local states. The expression in Eq.~\eqref{eq:PuritySimp} corresponds to repeating this many times over different initial states and recording the average result. Compared to other indicators of irregular or chaotic dynamics used in the discrete space-time setting (e.g.\ damage spreading), this quantity involves more replicas of the system and it is natural to expect it to provide a more stringent criterion for chaos.

Let us now provide a quantitative characterisation of LOE growth. First we note that permutation gates (cf.\ Eq.~\eqref{eq:permutationproperty}) for $q=2$ are part of the Clifford group~\cite{szaszschagrin2025entanglement, feng2025dynamics} and therefore map products of Pauli matrices into products of Pauli matrices~\cite{nielsen2000quantum}. While this property can be directly verified since there are only $24$ permutation matrices for $q=2$, it can be intuitively understood by noting that, for $q=2$, permutations do not implement universal classical computation. In contrast, they do so for $q>2$ as they can also implement Toffoli gates. The Clifford property means that at all times $\ket{\mathcal{O}_x(t)}$ is the superposition of at most three product states and thus the LOE is bounded by a constant, which excludes chaotic behaviour. 
Notably, this lack of chaoticity for $q=2$ is missed by damage spreading and OTOCs, which show the same behaviour for all $q\geq2$~\cite{bertini2025quantum}. Indeed, damage spreading is the classical analogue of OTOCs. As such, it signals ``classical scrambling" rather than chaos.

To make progress for $q>2$, we consider the case where 
$U(x,\tau)$ are random permutation gates (i.e.\ random unitary matrices fulfilling Eq.~\eqref{eq:permutationproperty}) independently distributed for each $x$ and $\tau$~\cite{bertini2025quantum}. In this case we make the customary approximation of considering the annealed average of Eq.~\eqref{eq:LOE}, i.e., we bring the average inside the logarithm and focus on the average purity given by 
\begin{equation}
\label{eq:puritydiagram}
  \mkern-8mu
  \overline{\mathcal{P}}_{\mathcal{O}}(t)=
  q^{2L} 
  \begin{tikzpicture}[baseline={([yshift=-0.6ex]current bounding box.center)},scale=0.45]
    \nctgridLine{-4.5}{0.5}{-4.7}{0.7}
    \foreach \x in {-3,-1,...,8}{\prop{\x}{1}{colPerm}{4}}
    \foreach \x in {-4,-2,...,7}{\prop{\x}{2}{colPerm}{4}}
    \foreach \x in {-3,-1,...,8}{\prop{\x}{3}{colPerm}{4}}
    \foreach \x in {-4,-2,...,7}{\prop{\x}{4}{colPerm}{4}}
    \foreach \x in {-5,...,3}{\circle{\x+0.5}{4.5}}
    \foreach \x in {4,...,6}{\square{\x+0.5}{4.5}}
    \foreach \x in {-4,-2,...,7}{\flatLD{\x+0.5}{0.5}}
    \foreach \x in {-5,-3,...,6}{\flatRD{\x+0.5}{0.5}}
    \node at (1.75,0.125) {\scalebox{0.8}{${}^{\mathcal{O}}$}};
    \draw[gray,decorate,decoration={brace,amplitude=3pt}] 
    (0.5,0.25)--(-4.5,0.25) node[midway,below,yshift=-2pt]{$L$};
    \draw[gray,decorate,decoration={brace,amplitude=3pt}] 
    (6.5,0.25)--(2.5,0.25) node[midway,below,yshift=-2pt]{$L-1$};
    \draw[gray,decorate,decoration={brace,amplitude=3pt}] (-4.5,4.75) -- (3.5,4.75)
    node[midway,above] {$L+2\ell$};
    \draw[gray,decorate,decoration={brace,amplitude=3pt}] (4.5,4.75) -- (6.5,4.75)
    node[midway,above] {$L-2\ell$};
    \draw[semithick,gray,|<->|] (8,0.5) -- (8,4.5) node[midway,right,yshift = 10pt] {$2t$};
  \end{tikzpicture}\mkern-24mu,
  \mkern-8mu
\end{equation}
with the averaged local gate
\begin{equation} \label{eq:defAveraged}
   \begin{tikzpicture}[baseline={([yshift=-0.6ex]current bounding box.center)},scale=0.65]
    \prop{0}{0}{colPerm}{2k}
  \end{tikzpicture}\coloneqq
  \overline{U(x,\tau)^{\otimes_{\rm r} 2k}} \eqqcolon  \bar{U}_{2k}.
\end{equation}
The top and bottom boundary states in Eq.~\eqref{eq:puritydiagram} are the same as in Eq.~\eqref{eq:PuritySimp}, and we assume periodic boundary conditions in space (i.e.\ the open legs on the left are connected to those on the right). For $q\ge 2k$, the average over random permutations reduces the number of states on each leg from $q^{2k}$ to $B_{2k}$, where the latter is the \emph{Bell number}~\cite{bell1934exponential,rota1964number} fulfilling $B_0=B_1=1$, and $B_n=\sum_{i=0}^{n-1}\binom{n-1}{i} B_i$. These states correspond one-to-one to partitions of a set of $2k$ elements, therefore we will refer to them as \emph{partition states} and denote them by $\{\ket*{\Pi_{j}}\}_{j=1}^{B_{2k}}$. More precisely, the averaged local gate $\bar{U}_{2k}$ becomes a projector to the subspace $\mathrm{Span}\big\{\ket*{\Pi_{j}}\otimes\ket*{\Pi_{j}}\big\}_{j=1}^{B_{2k}}$~\cite{Note20}. For instance, in the case of $2k=8$, the number of partition states is $B_8=4140$, and include $\ket*{\circleSA^{(8)}}$, $\ket*{\squareSA^{(8)}}$, $\ket*{\triangleSA^{(8)}}$, while for $2k=4$ we have $B_{4}=15$ partition states, including $\ket*{\flatSA^{(4)}}$, $\ket*{\circleSA^{(4)}}$ and $\ket*{\squareSA^{(4)}}$.

The unitarity of the local (non-averaged) gate 
$U_{2k}=U(x,t)^{\,\otimesR\,2k}$ implies $\bra*{\circleSA^{(4)}_{2}} U_4 = \bra*{\circleSA^{(4)}_{2}}$ and $\bra*{\squareSA^{(4)}_{2}} U_4 = \bra*{\squareSA^{(4)}_{2}}$, while the permutation property of the gates  (cf.\ Eq.~\eqref{eq:permutationproperty}) gives $U_k\ket*{\flatSA^{(4)}_{2}} = \ket*{\flatSA^{(4)}_{2}}$. Since these properties survive the average, for $L\geq t$ we can reduce Eq.~\eqref{eq:puritydiagram} to
\begin{equation}
\label{eq:purityoverlap}
  \bar{\mathcal{P}}_{\mathcal{O}}(t)
  =
  q^{2t+2\ell} 
  \Big[\mkern-4mu \bra*{\circleSA^{(4)}_{4\ell+1}\!}\otimes\!
  \bra*{\!\phi_{t-1-\ell}\!}\otimes\!
  \bra*{\squareSA^{(4)}_{1}\!}\mkern-4mu\Big]\mkern-4mu
  \ket*{\psi_{t+\ell}\!}, 
\end{equation}
where we for concreteness assumed $L$, $2\ell$ to be even, and introduced the states
\begin{equation}
\label{eq:psiphin}
\hspace{-.25cm}\bra{\phi_n} =   \hspace{-.25cm}\begin{tikzpicture}[baseline={([yshift=-0.6ex]current bounding box.center)},scale=0.5]
    \foreach \x in {1}{\prop{\x}{1}{colPerm}{4}}
    \foreach \x in {0,2}{\prop{\x}{0}{colPerm}{4}}
    \foreach \x in {-1,1,3}{\prop{\x}{-1}{colPerm}{4}}
    \foreach \x in {1,0,...,-1}{\square{2.5-\x}{0.5+\x}}
    \foreach \x in {1,0,...,-1}{\circle{-.5+\x}{0.5+\x}}
    \draw[gray,decorate,decoration={brace,amplitude=5pt}] (1.5,1.75) -- (1.5+2.25,1.75-2.25) node[midway,xshift=6.5pt,yshift=6.5pt] {$n$};
  \end{tikzpicture}, \quad 
\ket{\psi_n} =   \hspace{-.25cm}\begin{tikzpicture}[baseline={([yshift=-0.6ex]current bounding box.center)},scale=0.5]
    \foreach \x in {1}{\prop{\x}{1}{colPerm}{4}}
    \foreach \x in {0,2}{\prop{\x}{2}{colPerm}{4}}
    \foreach \x in {-1,1,3}{\prop{\x}{3}{colPerm}{4}}
    \node at (1.625,0.125) {\scalebox{0.8}{${}^{\mathcal{O}}$}};
    \foreach \x in {0,...,2}{\flatLD{0.5-\x}{0.5+\x}}
    \foreach \x in {0,...,2}{\flatRD{1.5+\x}{0.5+\x}}
    \draw[gray,decorate,decoration={brace,amplitude=5pt}] (0.675,0.325) -- (-1.675,2.675) node[midway,xshift=-6.5pt,yshift=-6.5pt] {$n$};
  \end{tikzpicture}. 
\end{equation}
In this language, the problem of computing LOE is rephrased as that of estimating an overlap between two vectors in a chain of $2t+2\ell$ qudits with $B_4=15$ internal states, which, in line with our physical interpretation of Eq.~\eqref{eq:PuritySimp}, we expect to be increasingly more suppressed as the chain length increases. This process can be characterised exactly in the limit of large $q$.  

\prlsection{Large $q$ expansion}The states $\bra{\phi_n}$ and $\ket{\psi_n}$ from Eq.~\eqref{eq:psiphin} can be conveniently defined inductively by introducing the rectangular matrices
\begin{equation}
  \begin{aligned}
    M_m &=  \begin{tikzpicture}[baseline={([yshift=-2.2ex]current bounding box.center)},scale=0.5]
      \foreach \x in {-3,-1,1,3,5}{\prop{\x}{5}{colPerm}{4}}
      \foreach \x in {4}{\flatLD{0.5-\x}{0.5+\x}}
      \foreach \x in {4}{\flatRD{1.5+\x}{0.5+\x}}
      \draw[gray,decorate,decoration={brace,amplitude=5pt}] (-3.625,5.5) -- (5.625,5.5) node[midway,above,yshift=2pt] {$2m$};
    \end{tikzpicture},\\
    N_m &=  \begin{tikzpicture}[baseline={([yshift=0.9ex]current bounding box.center)},scale=0.5]
      \foreach \x in {-3,-1,1,3,5}{\prop{\x}{5}{colPerm}{4}}
      \foreach \x in {4}{\circle{0.5-\x}{1.5+\x}}
      \foreach \x in {4}{\square{1.5+\x}{1.5+\x}}
      \draw[gray,decorate,decoration={brace,amplitude=5pt}] (5.625,4.5) -- (-3.625,4.5) node[midway,below,yshift=-1.5pt] {$2m$};
    \end{tikzpicture}.
    \label{eq:matNn}
  \end{aligned}
\end{equation}
In terms of the latter we have~\footnote{The form of $\ket{\psi_1}$ depends on the specific operator $\mathcal{O}$ under consideration. For definiteness, in the following we take it to be real and have diagonal elements equal to $\pm 1$. Our treatment, however, can be easily carried out for any initial operator in diagonal form.}
\begin{equation} \label{eq:evoleq}
  \begin{aligned}
    \ket{\psi_n} & = M_n \ket{\psi_{n-1}}, & \ket{\psi_1}& = \bar U_4 
    \ket{\blackflatSA{\mathcal{O}}_{4} \flatSA_4},\\
    \bra{\phi_n} & =  \bra{\phi_{n-1}} N_n, & 
    \bra{\phi_1}& = \bra{\circleSA_{4} \squareSA_4}  \bar U_4.
  \end{aligned}
\end{equation}
For the initial conditions of interest here, the action of $N_n$ ($M_n$) reduces the norm of $\bra{\phi_n}$ ($\ket{\psi_n}$) at each application and the idea is to find the contributions that are suppressed the least. In the case of $\bra{\phi_n}$ this procedure can be straightforwardly implemented  to give~\cite{Note20}
\begin{equation}
\label{eq:phileadingorder}
\bra{\phi_n} = q^{-n}
\smashoperator{\sum_{\substack{\ell_1,\ell_2,\ell_3\geq 0 \\ \ell_1+\ell_2+\ell_3 =n}}}
B_{\ell_1, \ell_2, \ell_3 }
  \prescript{(3,1,6)\mkern-8mu}{}{\bra*{ \ell_1, \ell_2, \ell_3}}
+ o\Big(\frac{1}{q^n}\Big),
\end{equation}
where $f(x)=o(x)$ is little-o notation and means $\lim_{x\to0}f(x)/x=0$, the $q$-independent  coefficients $B_{\ell_1, \ell_2, \ell_3 }$ are determined recursively, and we introduced the states
\begin{equation}\label{eq:defStLC}
  \mkern-8mu
  \ket{\ell_1,\ell_2,\ldots,\ell_n}^{\underline{c}}
  =\ket*{\Pi_{c_1,2\ell_1} \Pi_{c_2,2\ell_2} \cdots \Pi_{c_n,2\ell_n}}\!,
  \mkern-8mu
\end{equation}
involving $n$ different domains of partition states determined by the $n$-tuple
$\underline{c}=(c_1,c_2,\ldots,c_n)$. 

\begin{figure}
\includegraphics[width=\columnwidth]{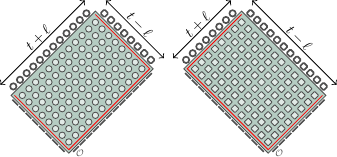}
  \caption{\label{fig:figMem} 
  Schematic illustration of the dominant contribution to local operator purity in cases with $\ell>0$ and $\ell<0$. The large rectangle represents (up to normalisation) the partition sum from Eq.~\eqref{eq:puritydiagram} after taking into account the unitarity and locality of the interaction, and the symbols outside the border represent the boundary conditions given by the partition states. Interpreting purity as a partition sum of a two-dimensional statistical spin model, the leading contribution comes from the bulk configuration that minimises the lengths of domain walls (red lines) between different partition states~\cite{jonay2018coarse,zhou2020entanglement}. Note that for $\ell=0$, both the configurations above give the same contribution, which explains the prefactor two in Eq.~\eqref{eq:puritymain}.
  }
\end{figure}

In the case of $\ket{\psi_n}$ keeping track of the leading-order contribution is complicated by a large number of cancellations (due to the orthogonality between $\ket{{\flatSA_{4}}}$ and $\ket*{\blackflatSA{\mathcal{O}}_{4}}$).
To make progress, we note that there exist subspaces of partition states that are invariant under the action of $M_{n}$. We start by introducing the following subspaces of partition states on a chain of length $n$
\begin{equation}
  \mkern-4mu
  \begin{aligned}
    \mathcal S_{1,n} &=
    \mathrm{Span}\mkern-2mu
    \big\{\mkern-4mu\ket*{\Pi_{i_1}\ldots\Pi_{i_{n}}}\mkern-4mu;\mkern-4mu&
    i_j&\!\in\!\{2,9,10,11,15\}\mkern-4mu\big\},\\
    \mathcal S_{2,n}&=\mathrm{Span}\mkern-2mu
    \big\{\mkern-4mu\ket*{\Pi_{i_1}\ldots\Pi_{i_{n}}}\mkern-4mu;\mkern-4mu&
    i_j&\!\in\!\{4,9,13,14,15\}\mkern-4mu\big\},\\
    \mathcal S_{3,n}&=\mathrm{Span}\mkern-2mu
    \big\{\mkern-4mu\ket*{\Pi_{i_1}\ldots\Pi_{i_{n}}}\mkern-4mu;\mkern-4mu&
    i_j&\!\in\!\{7,11,12,14,15\}\mkern-4mu\big\},\\
    \mathcal S_{4,n}&=\mathrm{Span}\mkern-2mu
    \big\{\mkern-4mu\ket*{\Pi_{i_1}\ldots\Pi_{i_{n}}}\mkern-4mu;\mkern-4mu&
    i_j&\!\in\!\{5,10,12,13,15\}\mkern-4mu\big\}.
  \end{aligned}
  \mkern-4mu
\end{equation}
As shown in SM~\cite{Note20}, the special feature of these states is that they can be factorised in replica space: they can all be written in terms of two factors, one of which is $\ket*{\flatSA^{(1)}_{n}}$. E.g., the states in $\mathcal S_{1,n}$ can all be written as $\ket*{\flatSA^{(1)}_{n}}\otimesR\ket*{\phi^{(3)}_{n}}$. This implies~\cite{Note20}
\begin{restatable}{property}{firstproperty}
  \label{prop:p1}
  For every $\ket{\phi}$ in $\mathcal{S}_{2n}$, where $\mathcal S_{n}\equiv\bigoplus_{j=1}^4\mathcal S_{j,n}$, and $\mathcal{S}_{n}^{\ast}$ is its dual space, one has  
  \begin{equation} \label{eq:prop1Eq}
    M_n \ket{\phi} \in 
    \mathcal S_{2n+2}, \qquad
    \bra{\phi} M_{n+1} \in \mathcal S^{\ast}_{2n-2}.
  \end{equation}
\end{restatable}
\noindent
In other words, the evolution defined by $M_n$ is closed for states in $\mathcal S_{2n}$. Moreover, recalling that $\braket*{\flatSA_{4}}{\blackflatSA{\mathcal{O}}_4}=0$, Property~\ref{prop:p1} also implies~\cite{Note20}
\begin{restatable}{property}{secondproperty}
  \label{prop:p2}
  For every $\ket{\phi}\in\mathcal S_{2n}$ we have $\braket{\phi}{\psi_n}=0$.
\end{restatable}
These two properties allow us to find the leading order contribution to $\ket{\psi_n}$ by focusing on the components of $\ket{\psi_n}$ that are \emph{not} in $\mathcal{S}_{2n}$.
At each time step we decompose the state as $\ket{\psi_n}  = \ket{\psi_n}' + \ket{\psi_n}''$, where $\ket{\psi_n}'' \in \mathcal S_{2n}$,
and $\ket{\psi_n}^{\prime}\notin \mathcal{S}_{2n}$.
Then, Property~\ref{prop:p1} implies $\ket{\psi_n}^{\prime}=\left(M_{n} \ket{\psi_{n-1}}^{\prime}\right)^{\prime}$, as $M_n\psi_{n-1}^{\prime\prime}$ has no component out of $\mathcal{S}_{2n}$. Using this observation, we are able to consistently determine the leading-order contribution to $\ket{\psi_n}^{\prime}$, while the orthogonality condition in Property~\ref{prop:p2} implies that $\ket{\psi_n}^{\prime\prime}$ gives subleading corrections~\cite{Note20}. 
Finally, we find that the leading order of $\ket{\psi_n}$ takes a form similar to Eq.~\eqref{eq:phileadingorder}, but the coefficients decay twice as fast---as $q^{-2n}$---and the relevant states (cf.\ \eqref{eq:defStLC}) are richer. They are given by $18$ different configurations of partition states $\underline{c}=(c_1,c_2,c_3,c_4,c_5)$ with (up to) $4$ domain walls~\cite{Note20}.

\prlsection{Results} Plugging the leading-order forms of $\bra{\phi_n}$ and $\ket{\psi_n}$ in Eq.~\eqref{eq:purityoverlap}, we find the following exact result~\cite{Note20}  
\begin{equation} \label{eq:puritymain}
  \bar{\mathcal{P}}_{\mathcal{O}}(t) 
  = \frac{1+\delta_{\ell,0}}{q^{t-\ell}} + o\!\left(\!\frac{1}{q^{t-\ell}}\!\right), 
\end{equation}
which is the main result of this letter. This equation shows that for large $q$, the LOE does indeed grow linearly in time for any finite $\ell$, demonstrating a chaotic behaviour of random permutation circuits. We note that while it is known that random permutation circuits with $q\geq 3$ form permutation $k$-designs in linear depth~\cite{gowers1996almost,chen2024incompressibility,gay2025pseudorandomness}, this does not imply linear growth of LOE at times $t<L$~\cite{jacoby2025long}. Our results apply precisely to the short time regime, $t<L$, showing linear growth of LOE and giving a precise prefactor.

Our result can be interpreted by generalising the entanglement membrane picture of Refs.~\cite{jonay2018coarse,zhou2020entanglement} to the random-permutation setting. As in the standard case, the idea is to interpret circuit-averaged multi-copy quantities as the partition function of a classical spin model, whose degrees of freedom are the partition states emerging from the averaging.
Notably, there are many more partition states than the `pairing states' occurring in random unitary circuits ($B_{2k}$ vs $k!$)~\cite{bertini2025quantum}. In this framework, the large-$q$ limit corresponds to the low-temperature limit of the partition function, which is dominated by the lowest energy configuration depicted in Fig.~\ref{fig:figMem}.
In a given partition state, uniform patches have zero energy and the energy cost only comes from the domain walls between different patches. At leading order, the line tension of different domain walls is given by the logarithm of the overlap between the neighbouring partition states, which are $\braket*{\circleSA^{(4)}}{\squareSA^{(4)}}=\braket*{\circleSA^{(4)}}{\flatSA^{(4)}}=\braket*{\squareSA^{(4)}}{\flatSA^{(4)}}=1/q$. Thus, Fig.~\ref{fig:figMem} indeed captures the leading contribution Eq.~\eqref{eq:puritymain}.

The membrane analysis can be generalized beyond diagonal operators, and can thus be employed to characterise the general expression in Eq.~\eqref{eq:PurityGeneral}. The dominant configurations are still those depicted in Fig.~\ref{fig:figMem}, where, however, $\circleSA^{(4)}$, $\squareSA^{(4)}$, and $\flatSA^{(4)}$ 
are replaced by $\circleSA^{(8)}$, $\squareSA^{(8)}$, and $\triangleSA^{(8)}$ respectively. Computing their overlaps we now have $\braket*{\triangleSA^{(8)}}{\circleSA^{(8)}}=\braket*{\circleSA^{(8)}}{\squareSA^{(8)}}=\braket*{\triangleSA^{(8)}}{\squareSA^{(8)}}=1/q^2$, and therefore we expect that for non-diagonal operators the averaged operator purity decays twice as fast (with the same asymptotic $q$ dependence of random unitary circuits, cf.~Ref.~\cite{zhou2019emergent}). 

\begin{figure}
    \centering
    \includegraphics[width=\linewidth]{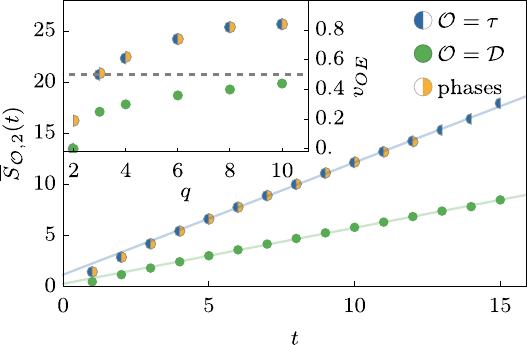}
    \caption{Averaged R\'enyi-$2$ local operator entanglement $\overline{S}_{\mathcal{O},2}(t)$ for $q=3$ for an off-diagonal operator $\tau$ (blue), a diagonal operator $\mathcal{D}$ (green), and $\tau$ under the dynamics with phases (yellow), and their linear fits. Note that for the off-diagonal operator $\overline{S}_{\mathcal{O},2}(t)$ grows approximately twice as fast as for the diagonal operator, ($v_{\mathrm{OE},\tau}=0.50$ vs $v_{\mathrm{OE},\mathcal{D}}=0.25$), and dynamics with phases is on top of the data without phases. The inset shows the numerically estimated velocities for various $q$. Note that for $q \gg 1$ and the diagonal operator the velocity approaches $1/2$, agreeing with the analytical results (dashed line). We also observe that at $q=2$, only dynamics with phases shows non-zero velocity, as it is not Clifford. The off-diagonal operator $\tau$ is the one-site cyclic shift of basis states, $\tau\ket{s}=\ket{s+1\pmod q}$, while the diagonal operator $\mathcal{D}$ is taken to be $\mathrm{diag}(+1,-1,+1,\ldots)$, and for odd $q$ we need to further normalise it as $\tr[\mathcal{D}]=0$ and $\tr[\mathcal{D}^2]=q$.
    }
    \label{fig:slope}
\end{figure}

To test these results in the finite-$q$ regime, we perform a numerical investigation exploiting the automaton nature of the dynamics~\cite{iaconis2019anomalous}. Namely, we randomly sample initial configurations and evolve each one classically. Despite being optimised for automaton circuits, this approach cannot compute LOE for very large times because it requires an exponentially large number of samples since the purity becomes exponentially small~\cite{Note20}.  Our results are summarised in Fig.~\ref{fig:slope}, which reports the (annealed) average of $S_{\mathcal{O},2}(t)$, as a function of time $t$. The finite-$t$ data shows convincing linear scaling allowing us to numerically estimate the speed of entanglement growth $v_{\mathrm{OE}} \equiv \lim_{t\to\infty} \overline{S}_{\mathcal{O},2}(t)/(\log(q^2)t)$ --- plotted in the inset. The results in the large $q$ limit for diagonal operators agree with Eq.~\eqref{eq:puritymain}, which predicts $v_{\mathrm{OE}}=1/2$, and also confirms that the entanglement velocity is twice as large for off-diagonal operators. Interestingly, this property appears to remain true also at finite $q$.

Finally, we note that one could also consider an immediate generalisation of the permutation dynamics by adding two-site random phases to dynamics by multiplying $U(x,t)$ with a randomly chosen diagonal phase matrix, which amounts to changing the r.h.s.\ in Eq.~\eqref{eq:permutationproperty} as $\ket{\pi_{x,\tau}(n)}\mapsto e^{i \phi_{x,\tau}(n)} \ket{\pi_{x,\tau}(n)}$. As we show in Fig.~\ref{fig:slope} (see also SM~\cite{Note20}), as long as $q>2$ the behaviour of the averaged purity is similar, and it takes exactly the same form for diagonal observables. The only qualitative difference occurs at $q=2$, where the two-site permutations with added phases no longer have Clifford property, and therefore can support linear growth of LOE.

\prlsection{Discussion} In this Letter, we characterised quantum chaos in random permutation circuits with nearest-neighbour interactions by means of local operator entanglement~\cite{prosen2007operator, prosen2007is}. We showed that while the latter is bounded for qubits, for qudits it grows linearly in the thermodynamic limit. This suggests that random permutations on qudits are quantum chaotic.

We also argued that local operator entanglement is a better diagnostic of classical chaos than damage spreading --- the latter only measures sensitivity to initial conditions --- and we proposed it as a unified tool for the detection of chaos in both the quantum and classical realms.

Our work opens many questions for future research. In our view, the two most pressing ones are: (i) Develop a comprehensive understanding of local operator entanglement in classical dynamical systems, determining how it is related to the standard concepts of chaos theory such as Lyapunov exponents and Kolmogorov-Sinai entropy~\cite{arnold1989ergodic, arnold2012geometrical, ott2002chaos, cornfeld2012ergodic}; (ii) explore the implications that the quantum chaotic nature of random permutation has on their quantum computational properties.

\begin{acknowledgments}
  We thank Toma\v z Prosen for useful discussions. We acknowledge financial support from the Royal Society through the University Research Fellowship No.\ 201101 (B.\ B.), the Leverhulme Trust through the Early Career Fellowship No.\ ECF-2022-324 (K.\ K.), the Novo Nordisk Foundation under grant numbers NNF22OC0071934 and NNF20OC0059939 (D.\ M.). P.\ K. is partly funded by THEQUCO as part of the Munich Quantum Valley, which is supported by the Bavarian state government with funds from the Hightech Agenda Bayern Plus. 
\end{acknowledgments}

\bibliography{./LOEbib}
\clearpage
\onecolumngrid
\newcounter{equationSM}
\newcounter{figureSM}
\newcounter{tableSM}
\stepcounter{equationSM}
\setcounter{equation}{0}
\setcounter{figure}{0}
\setcounter{table}{0}
\setcounter{section}{0}
\makeatletter
\renewcommand{\theequation}{S\,\arabic{equation}}
\renewcommand{\thefigure}{S\,\arabic{figure}}
\renewcommand{\thetable}{S\,\arabic{table}}

In this Supplemental Material, we report some complementary results from the main text. In particular 
\begin{itemize}
  \item In Sec.~\ref{sec:SuppReductionForDiag} we show how LOE R\'enyi-2 purity reduces to a $4$ replica quantity for diagonal observables.
  \item In Sec.~\ref{sec:SuppAverage} we provide additional details on the permutation-averaged multi-replica gates.

  \item In Sec.~\ref{sec:SuppPhases} we briefly discuss the dynamics with additional non-trivial phases.
  \item In Sec.~\ref{sec:SuppLargeQ} we do the leading order in the large-$q$ expansion of the LOE purity.
  \item In Sec.~\ref{sec:SuppNum} we provide additional details on the numerical methods used to produce the data in the main text. 
\end{itemize}

\section{Reduction of Eq.~(\ref{eq:PurityGeneral}) for diagonal operators}\label{sec:SuppReductionForDiag}
Since permutation dynamics preserves diagonal operators in the computational basis, the time-evolved diagonal operator acting non-trivially on the Hilbert space $\mathcal{H}_t$, can be interpreted as a vector from $\mathcal{H}_t$. This mapping reduces the number of degrees of freedom, and gives an expression equivalent to Eq.~\eqref{eq:PurityGeneral}, which now only involves $4$ replicas instead of $8$.

To see the reduction to Eq.~\eqref{eq:PuritySimp} explicitly, we define a rectangular matrix $\mathcal{D}_1$, which maps a one-site vector, to the vectorised version of the operator whose diagonal is given by the former,
\begin{equation}
  \mathcal{D}_1=\sum_{a=0}^{q-1} \ketbraR{aa}{a}. 
\end{equation}
Using now the fact that $U(x,t)$ is a permutation of computational basis elements, 
\begin{equation}
  \mathcal{D}_1\otimes \mathcal{D}_1 U(x,t)= 
  \left[U(x,t)\otimesR U^{\ast}(x,t)\right]\mathcal{D}_1\otimes \mathcal{D}_1
  = U_2(x,t) \mathcal{D}_1\otimes \mathcal{D}_1,\qquad
  U_{2k}(x,t)=\left[U(x,t)\otimesR U^{\ast}(x,t)\right]^{\,\otimesR\, k}.
\end{equation}
Similarly, defining 
\begin{equation}
  \mathcal{D}_k=\mathcal{D}_1^{\,\otimesR\,k},
\end{equation}
we have
\begin{equation}
  \mathcal{D}_4 \otimes \mathcal{D}_4 U_4(x,t) 
  = U_8(x,t) \mathcal{D}_4\otimes \mathcal{D}_4.
\end{equation}
Noticing now that we have (cf.\ \eqref{eq:defStates8} and~\eqref{eq:defStates4})
\begin{equation}
\label{eq:states}
    \ket*{\circleSA^{(8)}}=\mathcal{D}_4\ket*{\flatSA^{(4)}},\qquad
    \ket*{\blackcircleSA{8}^{(\mathcal{O})}}=\mathcal{D}_4\ket*{\flatSA_{\mathcal{O},4}},
\end{equation}
we can rewrite Eq.~\eqref{eq:PurityGeneral} as
\begin{equation}\label{eq:purityDiagSM1}
\mathcal{P}_{\mathcal{O}}(t)=q^{4L}
  \mel*{ \squareSA^{(8)}_{L+\ell} \triangleSA^{(8)}_{L-\ell}}
  {\mathcal{D}_4^{\otimes L}\mathcal{U}_4(t)}
  {\flatSA^{(4)}_{L-1} \mkern-4mu\blackflatSA{(4)}_{\mathcal{O}}\mkern-4mu
  \flatSA^{(4)}_{L}},
\end{equation}
where we note that now everything is written in terms of $4$-copy quantities, except for the left vectors. These can be locally understood as a one-site $8$-copy (row) vector, multiplied from the right by $\mathcal{D}_4$, which gives a $4$-copy vector. More precisely, we have
\begin{equation}
  \bra*{\squareSA^{(8)}}\mathcal{D}_4=\frac{1}{q}   \bra*{\circleSA^{(4)}},\qquad
  \bra*{\triangleSA^{(8)}}\mathcal{D}_4=\frac{1}{q} \bra*{\squareSA^{(4)}},
\end{equation}
which gives Eq.~\eqref{eq:PuritySimp} after being inserted into~\eqref{eq:purityDiagSM1}.

\section{Average over permutations and partition states}\label{sec:SuppAverage}
The averaged gate $\bar{U}_{m}$ in Eq.~\eqref{eq:defAveraged} is a projector to the set of invariant states~\cite{collins2022weingarten}, which are in one-to-one correspondence with \emph{partitions} of a set of $m$ elements. Let $\Pi^{(m)}_{j}$ be a partition of a set of $m$ elements labelled by $j$,
\begin{equation} \label{eq:partitionDefSupp}
  \Pi^{(m)}_{j}=\{\pi_{m,j}^{(1)},\pi_{m,j}^{(2)},\ldots \pi_{m,j}^{(n_{m,j})}\},\qquad
  \pi_{m,j}^{(l)}\cap \pi_{m,j}^{(l^{\prime})}=\varnothing,\qquad
  \bigcup_{l=1}^{n_{m,j}} \pi_{m,j}^{(l)}=\{1,2,\ldots,m\}.
\end{equation}
Then we can define the corresponding partition state $\ket*{\Pi^{(m)}_{j}}$ as a uniform superposition of all configurations of $m$ replicas so that the sites labelled by indices from the same subset $\pi_{m,j}^{(l)}$ are in the same state, namely,
\begin{equation}
  \Rbraket{a_1 a_2\ldots a_m}{\Pi^{(m)}_{j}}=
  \prod_{l=1}^{n_{m,j}}
  \left(\frac{1}{\sqrt{q}} \sum_{b=0}^{q-1}
  \smashoperator[r]{\prod_{k\in\pi_{m,j}^{(l)}}}\delta_{a_k,b}\right),
\end{equation}
where the prefactor ensures normalisation $\braket*{\Pi^{(m)}_{j}}{\Pi^{(m)}_{j}}=1$.

The invariant states for \emph{two-site} permutation gates are the tensor products of the above states: indeed, it is straightforward to see that any permutation gate leaves $\ket*{\Pi^{(m)}_{j}}\otimes \ket*{\Pi^{(m)}_{j}}$ invariant (for any $j$). It turns out that these states exhaust the set of all invariant states~\cite{collins2022weingarten}, and $\bar{U}_m$ is a projector to $\mathrm{Span}[\{\ket*{\Pi^{(m)}_{j}}\otimes\ket*{\Pi^{(m)}_{j}}\}_{j=1}^{B_m}]$, where $B_m$ is the number of partitions of the set of $m$ elements, known as the \emph{Bell number},
\begin{equation}
  B_m=\sum_{k=0}^{m-1}\binom{m-1}{k} B_k,\qquad B_0=B_1=1.
\end{equation}
The explicit form of $\bar{U}_m$ is therefore given as
\begin{equation}
  \bar{U}_{m}=\sum_{i,j=1}^{B_m} W^{(m)}_{i,j} 
  \left[\ket*{\Pi^{(m)}_{i}}\otimes\ket*{\Pi^{(m)}_{i}}\right]
  \left[\bra*{\Pi^{(m)}_{j}}\otimes\bra*{\Pi^{(m)}_{j}}\right],
\end{equation}
where $W^{(m)}$ is the Weingarten matrix, given as an inverse of the overlap matrix, 
\begin{equation}
  W^{(m)}=\left.G^{(m)}\right.^{-1},\qquad
  G^{(m)}_{i,j}=\braket*{\Pi^{(m)}_{i}}{\Pi^{(m)}_{j}}^2.
\end{equation}
Note that for $m\le q^2$, all the partition states are linearly independent, and the overlap matrix is invertible. In the case of $m> q^2$, however, $G^{(m)}$ becomes singular, in which case $W^{(m)}$ has to be taken to be the pseudoinverse of $G^{(m)}$.

\subsection{Partition states for $m=4$}
For future convenience, we report partition states for $m=4$,
\begin{equation}
  \begin{aligned}
    \ket{\Pi_1}&=\frac{1}{\sqrt{q}}\sum_{a=0}^{q-1} \ketR{aaaa}, &
    \ket{\Pi_2}&=\frac{1}{q}\sum_{a=0}^{q-1}\sum_{b=0}^{q-1} \ketR{abbb}, &
    \ket{\Pi_3}&=\frac{1}{q}\sum_{a=0}^{q-1}\sum_{b=0}^{q-1} \ketR{aabb},\\
    \ket{\Pi_4}&=\frac{1}{q}\sum_{a=0}^{q-1}\sum_{b=0}^{q-1} \ketR{abaa}, &
    \ket{\Pi_5}&=\frac{1}{q}\sum_{a=0}^{q-1}\sum_{b=0}^{q-1} \ketR{aaab}, &
    \ket{\Pi_6}&=\frac{1}{q}\sum_{a=0}^{q-1}\sum_{b=0}^{q-1} \ketR{abba},\\
    \ket{\Pi_7}&=\frac{1}{q}\sum_{a=0}^{q-1}\sum_{b=0}^{q-1} \ketR{aaba}, &
    \ket{\Pi_8}&=\frac{1}{q}\sum_{a=0}^{q-1}\sum_{b=0}^{q-1} \ketR{abab}, &
    \ket{\Pi_9}&=\frac{1}{q^{\frac{3}{2}}}\sum_{a=0}^{q-1}\sum_{b=0}^{q-1}\sum_{c=0}^{q-1}
    \ketR{abcc},\\
    \ket{\Pi_{10}}&=\frac{1}{q^{\frac{3}{2}}}\sum_{a=0}^{q-1}\sum_{b=0}^{q-1}\sum_{c=0}^{q-1}
    \ketR{abbc},&
    \ket{\Pi_{11}}&=\frac{1}{q^{\frac{3}{2}}}\sum_{a=0}^{q-1}\sum_{b=0}^{q-1}\sum_{c=0}^{q-1}
    \ketR{abcb},&
    \ket{\Pi_{12}}&=\frac{1}{q^{\frac{3}{2}}}\sum_{a=0}^{q-1}\sum_{b=0}^{q-1}\sum_{c=0}^{q-1}
    \ketR{aabc},\\
    \ket{\Pi_{13}}&=\frac{1}{q^{\frac{3}{2}}}\sum_{a=0}^{q-1}\sum_{b=0}^{q-1}\sum_{c=0}^{q-1}
    \ketR{abac},\quad&
    \ket{\Pi_{14}}&=\frac{1}{q^{\frac{3}{2}}}\sum_{a=0}^{q-1}\sum_{b=0}^{q-1}\sum_{c=0}^{q-1}
    \ketR{abca},\quad&
    \ket{\Pi_{15}}&=\frac{1}{q^2}\sum_{a=0}^{q-1}\sum_{b=0}^{q-1}\sum_{c=0}^{q-1}\sum_{d=0}^{q-1}
    \ketR{abcd}.
  \end{aligned}
\end{equation}
Note that for simplicity, we suppressed the explicit dependence on $m=4$, i.e.\ $\ket{\Pi_{j}}=\ket*{\Pi^{(4)}_{j}}$. We will sometimes use this convention when the value of $m=4$ is clear from the context.

\section{Permutations with phases}\label{sec:SuppPhases}
Let us now consider the local gate that acts not only as a permutation, but also multiplies each local configuration with a random phase, i.e.\ for a two-qudit state $\ket{n}=\ket{n_1n_2}$ (with $n=q n_1+n_2$) we have
\begin{equation}
  U(x,t) \ket{n} = \mathrm{e}^{\mathrm{i} \phi_{x,t}(n)}\ket{\pi_{x,t}(n)},\qquad
  \pi_{x,\tau}\in S_{q^2},\quad 0\le \phi_{x,t} (n) \le 2 \pi.
\end{equation}
The first observation we make is that for diagonal observables the discussion of Sec.~\ref{sec:SuppReductionForDiag} can be straightforwardly repeated, as we have
\begin{equation}
  \left[U(x,t)\otimesR U^{\ast}(x,t)\right]=
  \mathcal{D}_1 \otimes \mathcal{D}_1 \left. U(x,t)\right|_{\phi_{x,t}\to 0},
\end{equation}
which implies that the LOE purity is given by precisely the same expression as~\eqref{eq:purityDiagSM1}, upon replacing $\mathcal{U}_4(t)$ by $\left.\mathcal{U}_4(t)\right|_{\phi_{x,t}\to 0}$.

However, for generic $\mathcal{O}$ there is no immediate reason to expect the LOE purity of permutations with phases to behave the same as the one with no phases, as we have
\begin{equation}
  \mathcal{U}_{8}(t)\neq \left.\mathcal{U}_{8}(t)\right|_{\phi_{x,t}\to 0}.
\end{equation}
Also after averaging over all the possible choices of local permutations and phases, we do not expect the two averaged LOE purities to behave identically, as the local averaged gate is not the same. The averaged replicated gate $\bar{U}_{2n}$ is --- just like the $\phi_{x,t}=0$ counterpart --- the projector to the space of invariant states, but the invariant states represent a \emph{subset} of partition states. In particular, the invariant states are parametrised by partitions $\Pi_j^{(2n)}$, so that each of the subsets $\pi_{2n,j}^{(l)}$ (cf.\ \eqref{eq:partitionDefSupp}) contains an equal number of even and odd numbers. The number of invariant states now therefore grows slower than $B_{2n}$ and is
\begin{equation}
  \bar{B}_{2n}=\sum_{k=0}^{n-1} \binom{n}{k}\binom{n-1}{k} \bar{B}_{2k},
\end{equation}
which is still larger than $n!$ (i.e.\ the number of invariant states of a random unitary gate).

\section{Details on the large $q$ expansion}\label{sec:SuppLargeQ}
\subsection{Proof of Property~\ref{prop:p1}}

Here we provide a proof of Property~\ref{prop:p1}, which we restate for convenience 
\begin{restatable}{property}{firstproperty}
  \label{prop:Suppp1}
  For every $\ket{\phi}$ in $\mathcal{S}_{2n}$, 
  where 
  $\mathcal S_{n}\equiv\bigoplus_{j=1}^4\mathcal S_{j,n}$, and $\mathcal{S}_{n}^{\ast}$ is its dual space, one has  
  \begin{equation} \label{eq:prop1Eq}
    M_n \ket{\phi} \in 
    \mathcal S_{2n+2}, \qquad
    \bra{\phi} M_{n+1} \in \mathcal S^{\ast}_{2n-2}.
  \end{equation}
\end{restatable}
This property follows directly from the observation that for \emph{any} permutation gate $U$ (cf.\ Eq.~\eqref{eq:permutationproperty}) and state $\ket{\psi_3}$ living in the space of two qudits  replicated three times ($\simeq (\mathbb C^{q^2})^{\otimes 3}$) one has
\begin{equation}
  U^{\,\otimesR\, 4}
  \left[\ket{\psi_3}  \otimes_{\rm r} \ket{\flatSA_{1,2}}\right] 
  =  
  (U^{\,\otimesR\, 3}\ket{\psi_3}) 
  \otimesR \ket{\flatSA_{1,2}},
\end{equation}
because the flat sum of computational basis states is left invariant by all permutations. Averaging now over all possible permutations $U$ we have that $\ket{\psi_3}$ is projected in the space spanned by partition states of three replicas, i.e., 
\begin{equation}
  \{\ket*{\Pi^{(3)}_{j}\Pi^{(3)}_{j}}, \quad j=1,\dots,5\}, 
\end{equation}
with
\begin{equation}
  \begin{aligned}
    \ket*{\Pi^{(3)}_{1}}=&\frac{1}{\sqrt q}\sum_{a=0}^{q-1} \ketR{aaa}, &
    \ket*{\Pi^{(3)}_{2}}=&\frac{1}{q}\sum_{a,b=0}^{q-1} \ketR{aab},&
    \ket*{\Pi^{(3)}_{3}}=&\frac{1}{q}\sum_{a,b=0}^{q-1} \ketR{aba}, \\
    \ket*{\Pi^{(3)}_{4}}=&\frac{1}{q}\sum_{a,b=0}^{q-1} \ketR{baa}, &
    \ket*{\Pi^{(3)}_{5}}=&\frac{1}{q^{3/2}}\sum_{a,b,c=0}^{q-1} \ketR{abc}\,.
  \end{aligned}
\end{equation}
Noting that 
\begin{equation}
  \mathrm{Span}\{\ket*{\Pi^{(3)}_{j}\Pi^{(3)}_{j}}\otimesR \ket{\flatSA_{1,2}}, 
  \quad j=1,\dots,5\} = \mathcal S_{1,2},
\end{equation}
we then have 
\begin{equation}
  \bar U_4 (\ket{\psi_3}  \otimes_{\rm r} \ket{\flatSA_{1,2}}) \in (\mathcal S_{1,2}). 
\end{equation}
Introducing the operator $S_{\mathrm{r}}$ implementing a cyclic permutation among replicas and reasoning analogously we have 
\begin{equation}
  \bar U_4 S^{j} \ket{\psi_3}   \otimes_{\rm r} \ket{\flatSA_{1,2}} \in (\mathcal S_{j+1,2}), \qquad j=1,2,3,4.
\end{equation}
This immediately gives  
\begin{equation}
  \bar U_4 \ket{\phi}\in \mathcal S_{j,2}, \quad \forall \ket{\phi}\in \mathcal S_{j,2},
  \qquad  \bra{\phi} \bar U_4 \in \mathcal S^{\ast}_{j,2},
  \quad \forall \bra{\phi}\in \mathcal S^{\ast}_{j,2}, 
\end{equation}
where the second equation is obtained by taking the Hermitian conjugation and noting $\bar U_4 = \bar U_4^\dag$. Hence we have that for any state $\ket{\phi}\in\mathcal S_{2}$ 
\begin{equation}
  \bar U_4 \ket{\phi}\in \mathcal S_{2}, \qquad 
  \bra{\phi} \bar U_4 \in \mathcal{S}_{2}^{\ast}.
\end{equation}
Finally, the left of Eq.~\eqref{eq:prop1Eq} directly follows from the fact that $M_n \ket{\phi}$ is written as a tensor product of $\bar U_4$ applied onto states in $ \mathcal S_{2}$. Similarly, the right of Eq.~\eqref{eq:prop1Eq} follows from the fact that projecting the first and last component of a state in $\mathcal S_{n}$ onto $\ket{\flatSA_{4,1}}$ gives a state in $\mathcal S_{n-2}$. 

\subsection{Proof of Property~\ref{prop:p2}}
Here we provide a proof of Property~\ref{prop:p2}, which we restate for convenience 
\begin{restatable}{property}{secondproperty}
  \label{prop:Suppp2}
  For every $\ket{\phi}\in\mathcal S_{2n}$ we have $\braket{\phi}{\psi_n}=0$.
\end{restatable}
This property follows immediately by observing that 
\begin{equation}
  \braket{\phi}{\psi_n} = \mel{\phi}{M_n \ldots M_2 \bar U_4}
  {\blackflatSA{\mathcal{O}}_4 \flatSA_4}, 
\end{equation}
and that, by virtue of Property~\ref{prop:p1}, if $\bra{\phi}\in \mathcal S_{2n}^\ast$, $\bra{\phi}M_n \ldots M_2 \bar U_4\in \mathcal S_{2}^\ast$. Indeed, at this point one simply needs to notice that vectors in $\mathcal S_{1}^\ast$ are orthogonal to $\ket{\blackflatSA{\mathcal{O}}_{4}}$. 
This follows directly from
\begin{equation}
  \ket{\flatSA_{4}}=\ket{\flatSA_{1}}^{\,\otimesR\, 4},\qquad
  \mel{\flatSA_{1}}{\mathcal{O}}{\flatSA_{1}}=
  \mel{\flatSA_{1}}{\mathcal{O}^{\ast}}{\flatSA_{1}}=0,
\end{equation}
and the observation that all vectors in $\mathcal{S}_2$ have at least one replica in the state $\ket{\flatSA_1}$.

\subsection{Leading order expression of $\bra{\phi_n}$}
\label{app:leadingphi}
To find the leading order contribution to $\bra{\phi_n}$ we begin by noting 
\begin{equation}
  \label{eq:aboverel}
  \begin{aligned}
    \bra{\Pi_3 \Pi_6} \bar U_4 &= \frac{1}{q} 
    \Big(\bra{\Pi_3 \Pi_3} +\bra{\Pi_6 \Pi_6} + \bra{\Pi_1 \Pi_1} \Big) 
    +o\big(\frac{1}{q}\big), \\
    \bra{\Pi_3 \Pi_1} \bar U_4 &=  \Big(\frac{1}{\sqrt q}+o\big(\frac{1}{q}\big)\Big) 
    \Big(\bra{\Pi_3 \Pi_3} +\bra{\Pi_1 \Pi_1} \Big),\\
    \bra{\Pi_1 \Pi_6} \bar U_4 &=  \Big(\frac{1}{\sqrt q}+o\big(\frac{1}{q}\big)\Big)
    \Big(\bra{\Pi_6 \Pi_6} +\bra{\Pi_1 \Pi_1} \Big),
  \end{aligned}
\end{equation}
where the $o({1}/{q})$ on the first line involve diagonal sums of partition states, i.e., terms of the form 
\begin{equation}
  \bra{\Pi_j \Pi_j},
\end{equation}
for some $j\in\{1,2,\ldots,15\}$. This means that at leading order we can write 
\begin{equation}
  \label{eq:topstate1}
  \bra{\phi_1}= \frac{1}{q} \Big(\bra{\Pi_3 \Pi_3}+\bra{\Pi_6 \Pi_6} + \bra{\Pi_1 \Pi_1} \Big)
  +o\big(\frac{1}{q}\big),
\end{equation}
where $o({1}/{q})$ is again a diagonal sum of partition states. We now claim that this expression can be extended to the $n$-th time step as follows 
\begin{equation}
  \label{eq:topstaten}
  \bra{\phi_n} = \frac{1}{q^n}
  \smashoperator[r]{\sum_{\substack{\ell_1,\ell_2,\ell_3\geq 0 \\ \ell_1+\ell_2+\ell_3 =n}}}
  B_{\ell_1, \ell_2, \ell_3} 
  \prescript{(3,1,6)\mkern-8mu}{}{\bra*{\ell_1, \ell_2, \ell_3}}
  +o\big(\frac{1}{q^n}\big).
\end{equation}
Here we introduced the states
\begin{equation}
  \prescript{(3,1,6)\mkern-8mu}{}{\bra*{\ell_1, \ell_2, \ell_3}}=
  \bra*{\underbrace{\Pi_{3}\cdots\Pi_3}_{2\ell_1}
  \underbrace{\Pi_{1}\cdots\Pi_1}_{2\ell_2}
  \underbrace{\Pi_{6}\cdots\Pi_6}_{2\ell_3}},
\end{equation}
and the coefficients $B_{\ell_1, \ell_2, \ell_3 }$ fulfilling  
\begin{equation}\label{eq:recursivetop}
  \begin{aligned}
    B_{\ell_1, \ell_2, \ell_3 } &=B_{\ell_1-1, \ell_2+1, \ell_3-1} 
    +  B_{\ell_1, \ell_2-1, \ell_3 }+B_{\ell_1-1, \ell_2, \ell_3 }
    +B_{\ell_1, \ell_2, \ell_3-1},\\
    B_{1,0,0} &= B_{0,1,0} = B_{0,0,1} = 1\,,\\
    B_{\ell_1, \ell_2,-1} &= B_{\ell_1, -1,\ell_3} = B_{-1,\ell_2, \ell_3} = 0.
  \end{aligned}
\end{equation}
Finally $o\big({1}/{q^n}\big)$ is a sum of terms of the form
\begin{equation} \label{eq:genericlittleo}
  \bra{\Pi_{k_1} \Pi_{k_1} \Pi_{k_2}\Pi_{k_2}\cdots \Pi_{k_n}\Pi_{k_n}},
  \qquad k_j\in\{1,2,\ldots,15\}.
\end{equation}
We prove the expression in Eq.~\eqref{eq:topstaten} by induction. First we note that it coincides with Eq.~\eqref{eq:topstate1} for $n=1$. Next we observe that, when $\bar U_4$ is applied to $\bra{\Pi_i \Pi_j}$ with $i\neq j$, it shrinks its norm by at least a factor $O(1/\sqrt q)$ (cf.\ Eq.~\eqref{eq:SuppMatU4} below). In fact, for $i=3$ the minimal shrinking is achieved for $j=1,9,12$, while for $i=6$ we have to choose $j=1,10,14$. Recalling Eq.~\eqref{eq:matNn} this means that when $N_n$ is applied to terms contained in $o\left({1}/{q^n}\right)$, it shrinks its norm by at least a factor $O(1/q)$, giving $o({1}/{q^{n+1}})$. Moreover, the resulting term is again a sum of terms like those in Eq.~\eqref{eq:genericlittleo} with $n\mapsto n+1$. Instead, applying $N_n$ to the first term in Eq.~\eqref{eq:topstaten} and using Eq.~\eqref{eq:aboverel} we have 
\begin{equation}
  \begin{aligned}
    &\begin{split}
      \frac{1}{q^{n+1}}
      \smashoperator[r]{\sum_{\substack{\ell_1,\ell_2,\ell_3\geq 0 \\ \ell_1+\ell_2+\ell_3 =n}}}
      B_{\ell_1, \ell_2, \ell_3} 
      \Big((1-\delta_{\ell_2,0})
      &\prescript{(3,1,6)\mkern-8mu}{}{\bra*{\ell_1+1, \ell_2-1, \ell_3+1}}
      +\prescript{(3,1,6)\mkern-8mu}{}{\bra*{ \ell_1+1, \ell_2, \ell_3}}\\[-1em]
      +&\prescript{(3,1,6)\mkern-8mu}{}{\bra*{ \ell_1, \ell_2+1, \ell_3}}
      +\prescript{(3,1,6)\mkern-8mu}{}{\bra*{ \ell_1, \ell_2, \ell_3+1}}\Big)
    \end{split}\\
    =&\frac{1}{q^{n+1}}
    \smashoperator[r]{\sum_{\substack{\ell_1,\ell_2,\ell_3\geq 0 \\ \ell_1+\ell_2+\ell_3 =n+1}}}
    (B_{\ell_1-1, \ell_2+1, \ell_3-1} +  B_{\ell_1, \ell_2-1, \ell_3 }
    + B_{\ell_1-1, \ell_2, \ell_3 }+  B_{\ell_1, \ell_2, \ell_3-1}) 
    \prescript{(3,1,6)\mkern-8mu}{}{\bra*{ \ell_1, \ell_2, \ell_3}}, 
  \end{aligned}
\end{equation}
where in the first line we used $\prescript{(c_1,c_2,c_3)\mkern-8mu}{}{\bra*{\ell_1,\ell_2,\ell_3}}=0$ if any of the $\ell_j$ take negative values. The resulting expression coincides with the sum in Eq.~\eqref{eq:topstate1} when evaluated at $n\to n+1$.

\subsection{Leading order expression of $\ket{\psi_n}$}
Let us now consider the state $\ket{\psi_n}$. We begin by noting that, for a local operator $\mathcal O$ of the form $\mathcal O={\rm diag}(\pm 1, \pm 1, \ldots, \pm 1)$, we can write the following exact expression
\begin{equation} \label{eq:psi1}
    \ket{\psi_1} = \alpha_1 \ket{\Pi_1\Pi_1}
    +\alpha_2\smashoperator{\sum_{j\in\{3,6,8\}}}\ket{\Pi_j \Pi_j} 
    +\alpha_3\smashoperator{\sum_{j\in\{2,4,5,7\}}}\ket{\Pi_j \Pi_j}
    + \alpha_4 \smashoperator{\sum_{j\in\{9,\ldots,14\}}}\ket{\Pi_j \Pi_j}
    + \alpha_5 \ket{\Pi_{15} \Pi_{15}},
\end{equation}
where we have
\begin{equation}
  \alpha_1=-\frac{2q}{A},\qquad
  \alpha_2 = \frac{q^2-2}{A},\qquad
  \alpha_3= \frac{2}{A},\qquad
  \alpha_4= -\frac{q}{A}, \qquad
  \alpha_5= \frac{3}{A},\qquad 
  A={(q-1)(q+2)(q^2-3)}.
\end{equation}

Next we observe that the last three terms in Eq.~\eqref{eq:psi1} belong to  $\mathcal S_{2}$ as defined in the main text. Therefore, because of Property~\ref{prop:p1}, they contribute to $\ket*{\psi_n}$ with a term in $\mathcal S_{2n}$. Removing their contribution and repeating this operation at each time step, $\ket*{\psi_n}$ can be split as follows
\begin{equation}
  \label{eq:splitcorrections}
  \ket*{\psi_n} = \ket*{\psi_n}'+ \ket*{\psi_n}'', 
\end{equation}
where $\ket*{\psi_n}''\in \mathcal S_{2n}$ and $\ket*{\psi_n}'\notin \mathcal S_{2n}$. Supposing now $\ket*{\psi'_n}=O(q^{-m})$, for some $m\geq0$, Property~\ref{prop:p2} gives  
\begin{equation}
  0=\prescript{\prime\prime\mkern-8mu}{}{\braket*{\psi_n}{\psi_n}}
  =\prescript{\prime\prime\mkern-8mu}{}{\braket*{\psi_n}{\psi_n}}^{\prime}
  + \|\ket*{\psi_n}''\|^2
  \quad \Rightarrow\quad  \|\ket*{\psi_n}''\| 
  =\sqrt{-\prescript{\prime\prime\mkern-8mu}{}{\braket*{\psi_n}{\psi_n}}^{\prime}}
  =O\big(\frac{1}{q^{m+\frac{1}{2}}}\big),
\end{equation}
where we used that the magnitude of the scalar product between two normalised states, of which one is in $\mathcal S_{2n}$ and one not, is at most $1/\sqrt{q}$. This means that $\ket*{\psi_n}''$ is subleading in comparison to $\ket*{\psi_n}'$. 

We now want to show that the leading order of $\ket*{\psi_n}$ (which is by the above argument the same as that of $\ket*{\psi_n^{\prime}}$) can be recast with the following ansatz 
\begin{equation} \label{eq:finalresultmain}
\ket{\psi_n}=\frac{1}{q^{2n}}
  \smashoperator{\sum_{\underline{c}\in \mathcal C}}
  \smashoperator[r]{\sum_{\substack{ \ell_1+\cdots+\ell_5 = n\\ \ell_j \geq 0}}}
  D^{\underline{c}}_{\ell_1, \ldots, \ell_5 } \ket{\ell_1, \ldots, \ell_5 }^{\underline{c}}
  + o\Big(\frac{1}{q^{2n}}\Big),
\end{equation}
if the coefficients $D^{\underline{c}}_{\ell_1,\ldots,\ell_5}$ satisfy an appropriate set of recurrence relations. Here $\mathcal{C}$ represents the set of all allowed configurations
\begin{equation}
  \mathcal{C}=\{\underline{p}_j,\underline{d}_j,\underline{e}_j\}_{j=1}^{6},
\end{equation}
given by 
\begin{equation}
  \begin{aligned}
    \underline{p}_1 &= (15,9,3,9,15), & \underline{p}_2 &= (15,12,3,9,15), & 
    \underline{p}_3 &= (15,9,3,12,15), & \underline{p}_4 &= (15,12,3,12,15), \\ 
    \underline{p}_5 &= (15,12,15,9,15), & \underline{p}_6 &= (15,9,15,12,15), \\ 
    \underline{d}_1 &= (15,10,6,10,15), & \underline{d}_2 &= (15,14,6,10,15), &
    \underline{d}_3 &= (15,10,6,14,15), & \underline{d}_4 &= (15,14,6,14,15), \\ 
    \underline{d}_5 &= (15,14,15,10,15), & \underline{d}_6 &= (15,10,15,14,15), \\ 
    \underline{e}_1 &= (15,11,8,11,15), & \underline{e}_2 &= (15,13,8,11,15), & 
    \underline{e}_3 &= (15,11,8,13,15), & \underline{e}_4 &= (15,13,8,13,15), \\ 
    \underline{e}_5 &= (15,13,15,11,15), & \underline{e}_6 &= (15,11,15,13,15). 
  \end{aligned}
\end{equation}

First we note that if we set $n=1$ and
\begin{equation} \label{eq:initialD} 
  \begin{aligned}
    D^{\underline{p}_j}_{0, 0, 1, 0, 0} &= D^{\underline{d}_j}_{0, 0, 1, 0, 0} 
    = D^{\underline{e}_j}_{0, 0, 1, 0, 0}  = \delta_{j,1}, & 
    D^{\underline{p}_j}_{1, 0, 0, 0, 0} &= D^{\underline{d}_j}_{1, 0, 0, 0, 0} 
    = D^{\underline{e}_j}_{1, 0, 0, 0, 0}  = 0 \\ 
    D^{\underline{p}_j}_{0, 1, 0, 0, 0} &= D^{\underline{d}_j}_{0, 1, 0, 0, 0} 
    = D^{\underline{e}_j}_{0, 1, 0, 0, 0}  = 0, & 
    D^{\underline{p}_j}_{0, 0, 0, 1, 0} &= D^{\underline{d}_j}_{0, 0, 0, 1, 0} 
    = D^{\underline{e}_j}_{0, 0, 0, 1, 0}  = 0 \\ 
    D^{\underline{p}_j}_{0, 0, 0, 0, 1} &= D^{\underline{d}_j}_{0, 0, 0, 0, 1} 
    = D^{\underline{e}_j}_{0, 0, 0, 0, 1}  = 0,  
  \end{aligned}
\end{equation}
then Eq.~\eqref{eq:finalresultmain} corresponds to the first line of Eq.~\eqref{eq:psi1}. To avoid double counting and to remove all the states in $\mathcal S_{2n}$, we additionally impose 
\begin{equation}
  \begin{aligned}
    D^{\underline{c}_1}_{\ell_1, \ell_2, 0, \ell_4, \ell_5}&=
    D^{\underline{c}_4}_{\ell_1, \ell_2, 0, \ell_4, \ell_5}
    =D^{\underline{c}_5}_{\ell_1, \ell_2, 0, \ell_4, \ell_5}
    =D^{\underline{c}_6}_{\ell_1, \ell_2, 0, \ell_4, \ell_5}=0,\\
    D^{\underline{c}_2}_{\ell_1, 0, \ell_3, \ell_4, \ell_5}&=
    D^{\underline{c}_2}_{\ell_1, \ell_2,0,0, \ell_5}
    =D^{\underline{c}_3}_{\ell_1, 0, 0, \ell_4, \ell_5}
    =D^{\underline{c}_3}_{\ell_1, \ell_2,\ell_3,0, \ell_5}
    =D^{\underline{c}_4}_{\ell_1, 0, \ell_3, \ell_4, \ell_5}
    =D^{\underline{c}_4}_{\ell_1, \ell_2,\ell_3,0, \ell_5}=0,\\
    D^{\underline{c}_5}_{\ell_1, 0, \ell_3, \ell_4, \ell_5}&=
    D^{\underline{c}_5}_{\ell_1, \ell_2,\ell_3,0, \ell_5}
    =D^{\underline{c}_6}_{\ell_1, 0, \ell_3, \ell_4, \ell_5}
    =D^{\underline{c}_6}_{\ell_1, \ell_2,\ell_3,0, \ell_5}=0, 
  \end{aligned}
\end{equation}
where $\underline{c}$ is a placeholder for the symbols $\{\underline{p},\underline{d},\underline{e}\}$. Note that these conditions are compatible with Eq.~\eqref{eq:initialD}. 

To express a recurrence relation between coefficients $D_{\ell_1,\ell_2,\ell_3,\ell_4,\ell_5}^{\underline{c}_j}$, we note the following relations
\begin{equation}\label{eq:SuppMatU4}
\begin{aligned}
  \bar U_4 \ket{\Pi_3 \Pi_{15}}  & = \frac{1}{q}\Big(\ket{\Pi_{3} \Pi_{3}}
  +\ket{\Pi_{15} \Pi_{15}}+\ket{\Pi_{9}\Pi_{9}}+\ket{\Pi_{12}\Pi_{12}}\Big) 
  +o\big(\frac{1}{q}\big),\\
  \bar U_4 \ket{\Pi_6 \Pi_{15}}  & = \frac{1}{q}\Big(\ket{\Pi_{6} \Pi_{6}}
  +\ket{\Pi_{15} \Pi_{15}}+\ket{\Pi_{10}\Pi_{10}}+\ket{\Pi_{14}\Pi_{14}}\Big) 
  +o\big(\frac{1}{q}\big),\\
  \bar U_4 \ket{\Pi_8 \Pi_{15}}  & = \frac{1}{q}\Big(\ket{\Pi_{8} \Pi_{8}}
  +\ket{\Pi_{15} \Pi_{15}}+\ket{\Pi_{11}\Pi_{11}}+\ket{\Pi_{13}\Pi_{13}}\Big) 
  +o\big(\frac{1}{q}\big),\\
  \bar U_4 \ket{\Pi_j \Pi_{15}}  & = \Big(\frac{1}{\sqrt q}+o\big(\frac{1}{q}\big)\Big)
  \Big(\ket{\Pi_{15} \Pi_{15}}+\ket{\Pi_{j}\Pi_{j}}\Big),\qquad j=9,\ldots,14,\\
  \bar U_4 \ket{\Pi_2 \Pi_{15}}  & = \frac{1}{q}\Big(\ket{\Pi_{2} \Pi_{2}}
  +\ket{\Pi_{15} \Pi_{15}}+\ket{\Pi_{9}\Pi_{9}}+\ket{\Pi_{10}\Pi_{10}}
  +\ket{\Pi_{11}\Pi_{11}}\Big) +o\big(\frac{1}{q}\big),\\
  \bar U_4 \ket{\Pi_4 \Pi_{15}}  & = \frac{1}{q}\Big(\ket{\Pi_{4} \Pi_{4}}
  +\ket{\Pi_{15} \Pi_{15}}+\ket{\Pi_{9}\Pi_{9}}+\ket{\Pi_{13}\Pi_{13}}
  +\ket{\Pi_{14}\Pi_{14}}\Big) +o\big(\frac{1}{q}\big),\\
  \bar U_4 \ket{\Pi_5 \Pi_{15}}  & = \frac{1}{q}\Big(\ket{\Pi_{5} \Pi_{5}}
  +\ket{\Pi_{15} \Pi_{15}}+\ket{\Pi_{10}\Pi_{10}}+\ket{\Pi_{12}\Pi_{12}}
  +\ket{\Pi_{13}\Pi_{13}}\Big) +o\big(\frac{1}{q}\big),\\
  \bar U_4 \ket{\Pi_7 \Pi_{15}}  & =\frac{1}{q}\Big(\ket{\Pi_{7} \Pi_{7}}
  +\ket{\Pi_{15} \Pi_{15}}+\ket{\Pi_{10}\Pi_{10}}+\ket{\Pi_{11}\Pi_{11}}
  +\ket{\Pi_{14}\Pi_{14}}\Big) + o\big(\frac{1}{q}\big),\\
  \bar U_4 \ket{\Pi_1 \Pi_{15}}  & = o\big(\frac{1}{q}\big).
\end{aligned}
\end{equation}
From them it follows that application of $M_n$ on the first term on the r.h.s.\ of Eq.~\eqref{eq:finalresultmain} gives
\begin{equation}
  \begin{aligned}
    &\frac{1}{q^{2n}}
    \smashoperator[l]{\sum_{\underline{c}\in \{\underline{p}, \underline{d}, \underline{e}\}}}
    \sum_{k=1}^6  
    \smashoperator[r]{\sum_{\substack{ \ell_1+\cdots+\ell_5 = n\\ \ell_j \geq 0}}}
    D^{\underline{c}_k}_{\ell_1, \ldots, \ell_5 } 
    \Big(\ket{\ell_1+1, \ell_2, \ell_3, \ell_4, \ell_5}^{\underline{c}_k}
    +\ket{\ell_1, \ell_2, \ell_3, \ell_4, \ell_5+1}^{\underline{c}_k}\\
    &+\sum _{j=1}^6 a_{k,j}(\ell_2) 
    \ket{\ell_1+1, \ell_2-1, \ell_3, \ell_4, \ell_5+1}^{\underline{c}_j}
    +\sum _{j=1}^6 b_{k,j}(\ell_4)  
    \ket{\ell_1+1, \ell_2, \ell_3, \ell_4-1, \ell_5+1}^{\underline{c}_j}\\
    &+\sum _{j=1}^6 c_{k,j}(\ell_2+1,\ell_4)  
    \ket{\ell_1+1, \ell_2-1, \ell_3+1, \ell_4-1, \ell_5+1}^{\underline{c}_j}
    +\sum_{j=1}^6 a_{j,k}(\ell_2+1) 
    \ket{\ell_1, \ell_2+1, \ell_3, \ell_4, \ell_5}^{\underline{c}_j} \\
    &+\sum_{j=1}^6 b_{j,k}(\ell_4+1)  
    \ket{\ell_1, \ell_2, \ell_3, \ell_4+1, \ell_5}^{\underline{c}_j}
    +\sum_{j=1}^6 c_{j,k}(\ell_2+1,\ell_4+1) 
    \ket{\ell_1, \ell_2+1, \ell_3-1, \ell_4+1, \ell_5}^{\underline{c}_j}\\
    &+\sum_{j=1}^6 h_{j,k}(\ell_2+1,\ell_4+1)  
    \ket{\ell_1, \ell_2+1, \ell_3, \ell_4-1, \ell_5+1}^{\underline{c}_j} 
    +\sum_{j=1}^6 g_{j,k}(\ell_2-1,\ell_4+1)  
    \ket{\ell_1+1, \ell_2-1, \ell_3, \ell_4+1, \ell_5}^{\underline{c}_j}\\
    &+\sum_{j=1}^6 l_{k,j}(\ell_3+1) 
    \ket{\ell_1, \ell_2, \ell_3+1, \ell_4, \ell_5}^{\underline{c}_j} 
    +\sum_{j=1}^6 l_{j,k}(\ell_3)  
    \ket{\ell_1+1, \ell_2, \ell_3-1, \ell_4, \ell_5+1}^{\underline{c}_j} \\
    &+\sum_{j=1}^6 m_{k,j}(\ell_4,\ell_3) 
    \ket{\ell_1, \ell_2, \ell_3+1, \ell_4-1, \ell_5+1}^{\underline{c}_j}
    +\sum_{j=1}^6 n_{k,j}(\ell_2,\ell_3) 
    \ket{\ell_1+1, \ell_2-1, \ell_3+1, \ell_4, \ell_5}^{\underline{c}_j}   \\
    &+\sum_{j=1}^6 o_{k,j}(\ell_2,\ell_3) 
    \ket{\ell_1, \ell_2+1, \ell_3-1, \ell_4, \ell_5+1}^{\underline{c}_j} 
    +\sum_{j=1}^6 p_{k,j}(\ell_4) 
    \ket{\ell_1+1, \ell_2, \ell_3-1, \ell_4+1, \ell_5}^{\underline{c}_j} \Big),
  \end{aligned}
\end{equation}
where, as in Sec.~\ref{app:leadingphi}, we used $\ket{ \ell_1, \ell_2, \ldots, \ell_n}^{\underline{c}}=0$ if any of the $\ell_j$ take negative values and introduced 
\begin{equation}
  \mkern-12mu
  \begin{aligned}
    a_{j,k}(x) &= \delta_{j,k}(1-\delta_{x,1}) + \delta_{x,1}\big((\delta _{j,1}
    +\delta _{j,2})\delta_{k,1}+(\delta_{j,3}+\delta_{j,4}) \delta_{k,3}\big),\\ 
    b_{j,k}(x) &= \delta_{j,k}(1-\delta_{x,1}) 
    + \delta_{x,1} \big((\delta _{j,1}+\delta _{j,3}) \delta_{k,1}
    +(\delta _{j,2}+\delta _{j,4}) \delta _{k,2}\big),\\
    c_{j,k}(x,y) &= \delta_{j,k}(1-\delta_{x,1})(1-\delta_{y,1}) 
    + (1-\delta_{x,1}) \delta_{y,1}  b_{j,k}(y) + (1-\delta_{y,1}) \delta_{x,1}  a_{j,k}(x) 
    + \delta_{x,1}\delta_{y,1}\delta_{k,1},\\
    g_{j,k}(x,y) &= \delta_{j,k}(1-\delta_{x,0})(1-\delta_{y,1}) 
    + (1-\delta_{x,0}) \delta_{y,1}  b_{j,k}(1) 
    + (1-\delta_{y,1}) \delta_{x,0}  a_{k,j}(1) 
    + \delta_{x,0}\delta_{y,1}(\delta_{j,1} + \delta_{j,3})(\delta_{k,1} + \delta_{k,2}),\\
    l_{j,k}(x) &= \delta_{j,k}(1-\delta_{x,1})+\delta_{x,1} 
    (\delta_{j,k}+ \delta _{k,2} \delta _{j,5}+\delta _{k,3} \delta _{j,6}),\\
    m_{j,k}(x,y) &= \delta_{j,k}(1-\delta_{x,1}) 
    + \delta_{y,0}(1-\delta_{x,1}) (\delta _{j,2} \delta _{k,5}+\delta _{j,3} \delta _{k,6}) 
    + \delta_{x,1}  b_{j,k}(1) ,\\
    n_{j,k}(x,y) &= \delta_{j,k}(1-\delta_{x,1}) 
    + \delta_{y,0}(1-\delta_{x,1}) (\delta _{j,2} \delta _{k,5}+\delta _{j,3} \delta _{k,6}) 
    + \delta_{x,1} a_{j,k}(1),\\
    o_{j,k}(x,y) &= \delta_{j,k}(1-\delta_{x,0}) + 
    \delta_{x,0} a_{k,j}(1)+ \delta_{y,1}(1-\delta_{x,0}) l_{j,k}(1),\\
    p_{j,k}(x,y) &= \delta_{j,k}(1-\delta_{x,0}) + \delta_{x,0}b_{k,j}(1) 
    + \delta_{y,1}(1-\delta_{x,0}) l_{j,k}(1).
  \end{aligned}
  \mkern-12mu
\end{equation}
This is compatible with Eq.~\eqref{eq:finalresultmain} if we require the coefficients $D^{\underline{c}_j}_{\ell_1, \ell_2, \ell_3, \ell_4, \ell_5}$ to fulfil the following recurrence condition 
\begin{equation}
  \begin{aligned}
    D^{\underline{c}_j}_{\ell_1, \ell_2, \ell_3, \ell_4, \ell_5} &= 
    D^{\underline{c}_j}_{\ell_1-1, \ell_2, \ell_3, \ell_4, \ell_5}
    +D^{\underline{c}_j}_{\ell_1, \ell_2, \ell_3, \ell_4, \ell_5-1}+
    \sum _{k=1}^6 a_{k,j}(\ell_2+1) 
    D^{\underline{c}_k}_{\ell_1-1, \ell_2+1, \ell_3, \ell_4, \ell_5-1}\\
    &+\sum _{k=1}^6 b_{k,j}(\ell_4+1)  
    D^{\underline{c}_k}_{\ell_1-1, \ell_2, \ell_3, \ell_4+1, \ell_5-1}
    +\sum _{k=1}^6 c_{k,j}(\ell_2+1,\ell_4+1) 
    D^{\underline{c}_k}_{\ell_1-1, \ell_2+1, \ell_3-1, \ell_4+1, \ell_5-1}\\
    &+\sum _{k=1}^6 a_{j,k}(\ell_2) 
    D^{\underline{c}_k}_{\ell_1, \ell_2-1, \ell_3, \ell_4, \ell_5}
    +\sum _{k=1}^6 b_{j,k}(\ell_4)  
    D^{\underline{c}_k}_{\ell_1, \ell_2, \ell_3, \ell_4-1, \ell_5}
    +\sum _{k=1}^6 c_{j,k}(\ell_2,\ell_4) 
    D^{\underline{c}_k}_{\ell_1, \ell_2-1, \ell_3+1, \ell_4-1, \ell_5}\\
    &+\sum _{k=1}^6 h_{j,k}(\ell_2,\ell_4) 
    D^{\underline{c}_k}_{\ell_1, \ell_2-1, \ell_3, \ell_4+1, \ell_5-1}
    +\sum _{k=1}^6 g_{j,k}(\ell_2,\ell_4)  
    D^{\underline{c}_k}_{\ell_1-1, \ell_2+1, \ell_3, \ell_4-1, \ell_5}\\
    &+\sum _{k=1}^6 l_{k,j}(\ell_3) 
    D^{\underline{c}_k}_{\ell_1, \ell_2, \ell_3-1, \ell_4, \ell_5}
    +\sum _{k=1}^6 l_{j,k}(\ell_3+1) 
    D^{\underline{c}_k}_{\ell_1-1, \ell_2, \ell_3+1, \ell_4, \ell_5-1} \\
    &+\sum _{k=1}^6 m_{k,j}(\ell_4+1,\ell_3-1)  
    D^{\underline{c}_k}_{\ell_1, \ell_2, \ell_3-1, \ell_4+1, \ell_5-1}
    +\sum _{k=1}^6 n_{k,j}(\ell_2+1,\ell_3-1) 
    D^{\underline{c}_k}_{\ell_1-1, \ell_2+1, \ell_3-1, \ell_4, \ell_5} \\
    &+\sum _{k=1}^6 o_{k,j}(\ell_2-1,\ell_3+1) 
    D^{\underline{c}_k}_{\ell_1, \ell_2-1, \ell_3+1, \ell_4, \ell_5-1}
    +\sum _{k=1}^6 p_{k,j}(\ell_4-1) 
    D^{\underline{c}_k}_{\ell_1-1, \ell_2, \ell_3+1, \ell_4-1, \ell_5},
  \end{aligned}
\end{equation}
with the conditions
\begin{equation}
  D^{\underline{c}_k}_{-1, \ell_2, \ell_3, \ell_4, \ell_5}
  =D^{\underline{c}_k}_{\ell_1, -1, \ell_3, \ell_4, \ell_5}
  =D^{\underline{c}_k}_{\ell_1, \ell_2, -1, \ell_4, \ell_5}
  =D^{\underline{c}_k}_{\ell_1, \ell_2, \ell_3, -1, \ell_5}
  =D^{\underline{c}_k}_{\ell_1, \ell_2, \ell_3, \ell_4, -1} = 0. 
\end{equation}
Finally, proceeding as in Sec.~\ref{app:leadingphi}, we one can directly show that the $o({1}/{q^{2n}})$ in Eq.~\eqref{eq:finalresultmain} becomes $o({1}/{q^{2n+2}})$ upon application of $M_n$ because all states that are suppressed less than $1/q^2$ are in $\mathcal S_{2n}$. This concludes the proof. 

\subsection{Purity at leading order}

To compute the purity we begin by observing that expanding both $\bra{\phi_{n}}$ and $\ket{\psi_{n}}$ in partition states one can rewrite Eq.~\eqref{eq:purityoverlap} as follows  
\begin{equation}
  \bar{\mathcal{P}}_{\mathcal{O}}(t) =  q^{2t+2\ell} 
  \sum_{\{k\}} \sum_{\{p\}} E_{\{k\}} F_{\{p\}} 
  \braket*{\underbrace{\Pi_3\cdots \Pi_3}_{4\ell+1} \Pi_{k_1}\Pi_{k_1}\Pi_{k_2}\Pi_{k_2} 
  \cdots \Pi_{k_{t-\ell-1}} \Pi_{k_{t-\ell-1}} \Pi_{6}}{\Pi_{p_{1}}\Pi_{p_{1}}\Pi_{p_{2}}\Pi_{p_{2}}\cdots \Pi_{p_{t+\ell}}\Pi_{p_{t+\ell}}}, 
\end{equation}
where we used the brickwork structure of the two states, and without loss of generality assumed $\ell\in\mathbb{N}_{0}$. Due to the staggered structure of bra and ket states, the overlaps appearing in the above equation are all upper bounded by $1/q$ which is attained for 
\begin{equation}
  \bra*{\underbrace{\Pi_{3}\Pi_{3}\cdots \Pi_{3}\Pi_{3}}_{4\ell+2j^*+1} 
  \Pi_{6}\Pi_{6} \cdots \Pi_{6}\Pi_{6}}, \quad 
  \ket*{\underbrace{\Pi_{3}\Pi_{3}\cdots \Pi_{3}\Pi_{3}}_{4\ell+2j^*+2} 
  \Pi_{6}\Pi_{6} \cdots \Pi_{6}\Pi_{6}}, \quad j^*=0,\ldots, t-\ell-1,
\end{equation}
and 
\begin{equation}
  \bra*{\underbrace{\Pi_{3}\Pi_{3}\cdots \Pi_{3}\Pi_{3}}_{4\ell+2j^*+1} 
  \Pi_{6}\Pi_{6} \cdots \Pi_{6}\Pi_{6}}, \quad
  \ket*{\underbrace{\Pi_{3}\Pi_{3}\cdots \Pi_{3}\Pi_{3}}_{4\ell+2j^*} 
  \Pi_{6}\Pi_{6} \cdots \Pi_{6}\Pi_{6}}, \quad j^*=0,\ldots, t-\ell-1. 
\end{equation}
Noting that all these bra states appear in the leading order expansion of $\bra{\phi_{n}}$ in Eq.~\eqref{eq:phileadingorder} and one of the ket states (two for $\ell=0$) appears in the leading order expansion of $\ket{\psi_{n}}$ in Eq.~\eqref{eq:finalresultmain} we have 
\begin{equation}
  \bar{\mathcal{P}}_{\mathcal{O}}(t) = 
  \frac{1}{q^{t-\ell}}
  \Big( B_{t-\ell-1, 0, 0} D^{\underline{p}_1}_{0,0, t+\ell,0, 0}
  + \delta_{\ell,0} B_{0,0,t-\ell-1} D^{\underline{d}_1}_{0,0, t+\ell,0, 0}\Big) 
  + o\big(\frac{1}{q^{t-\ell}}\big)\,.
\end{equation}
The expression in Eq.~\eqref{eq:puritymain} follows from noting $B_{m, 0, 0} =B_{0,0,m} =D^{\underline{p}_1}_{0,0,m,0, 0}=D^{\underline{d}_1}_{0,0,m,0,0}=1$, which follow directly from the recursive relations of the coefficients $B_{\ell_1,\ell_2,\ell_3}$ and $D^{\underline{c}_k}_{\ell_1, \ell_2, \ell_3, \ell_4, \ell_5}$.

\subsubsection{Different choices of parity of $L$ and $\ell$}\label{subsec:parityLl}
In obtaining the above expression, we assumed both $L$ and $2\ell$ to be even. For completeness let us just report the expressions for arbitrary parities of $L$ and $2\ell$. We start by noting that Eq.~\eqref{eq:purityoverlap} 
\begin{equation}
  \bar{\mathcal{P}}_{\mathcal{O}}(t)
  = \begin{cases}
    q^{2\lfloor t+ \ell+\frac{1}{2}\rfloor}
  \Big[\bra*{\circleSA^{(4)}_{4\ell+1}}\otimes
    \bra*{\phi_{\lfloor{t-(\ell+\frac{1}{2})\rfloor}}}\otimes
  \bra*{\squareSA^{(4)}_{1}}\Big]
    \ket*{\psi_{\lfloor t+\ell+\frac{1}{2}\rfloor}},& L \text{ even},\ \ell\ge 0,\\[0.5em]
    q^{2\lfloor t+ \ell\rfloor}
  \Big[\bra*{\circleSA^{(4)}_{4\ell-1}}\otimes
    \bra*{\phi_{\lfloor{t-\ell\rfloor}}}\otimes
  \bra*{\squareSA^{(4)}_{1}}\Big]
    \ket*{\psi_{\lfloor t+\ell\rfloor}},& L \text{ odd},\ \ell\ge\frac{1}{2},
  \end{cases}
\end{equation}
while for other values of $\ell\in\mathbb{Z}/2$ we use the left-right reflection symmetry of the local averaged gate to obtain 
\begin{equation}
  \left.\bar{\mathcal{P}}_{\mathcal{O}}(t)\right|_{\substack{L \text{ even}\\ \ell<0}} = 
  \left.\bar{\mathcal{P}}_{\mathcal{O}}(t)\right|_{\substack{L \text{ odd}\\ \ell\to\frac{1}{2}-\ell}},\qquad
  \left.\bar{\mathcal{P}}_{\mathcal{O}}(t)\right|_{\substack{L \text{ odd}\\ \ell\le 0}} = 
  \left.\bar{\mathcal{P}}_{\mathcal{O}}(t)\right|_{\substack{L \text{ even}\\ \ell\to\frac{1}{2}-\ell}}.
\end{equation}
Taking into account all the possible combinations of parity of $2\ell$ and $L$, as well as the sign of $\ell$, we can summarise the final result in a form identical to~\eqref{eq:puritymain}, upon replacing $\ell$ with its (appropriately shifted) absolute value, namely,
\begin{equation}
  \mathcal{P}_{\mathcal{O}}(t)=\frac{1+\delta_{\ell^{\ast},0}}{q^{t-\ell^{\ast}}}
  +o\Big(\frac{1}{q^{t-\ell^{\ast}}}\Big)
  ,\qquad
  \ell^{\ast}=
  \begin{cases}
    \lfloor |\ell| \rfloor, &\text{$L$ even},\\
    \lfloor |\ell-\frac{1}{2}| \rfloor,\quad &\text{$L$ odd}.
  \end{cases}
\end{equation}

\section{Details of numerical algorithms}\label{sec:SuppNum}
The backbone of the used numerical algorithms is that for a given realization of the permutation gates and a given initial computational basis  state $\ket{\underline{s}}$, the time evolution can be performed efficiently. Indeed, in this case the application of each local gate only modifies the string locally, which produces the time-evolved state $\ket{\underline{s}^t}$ with polynomial cost in $t$. Now we wish to use this fact to compute the averaged local operator purity $\bar{\mathcal{P}}_{\mathcal{O}}(t)$ in the most efficient way.

\subsection{Algorithm for general operators}
We start by expressing time-evolution of local operators in the computational basis, for which we need to setup some notation. Let $\ket{\underline{s}}=\ket*{s_0s_{\frac{1}{2}}\ldots s_{L-\frac{1}{2}}}$ be an initial configuration. Then $\ket{\underline{s}^{t}}=\mathcal{U}(t)\ket{\underline{s}}$ is the corresponding time-evolved state given by a \emph{single} computational-basis state.  Let now $\ket{\tilde{\underline{s}}_b}$ be a computational basis state, which is related to $\ket{\tilde{\underline{s}}_b}$ by replacing the site $L/2$  with the value $b$. Explicitly,
\begin{equation}
  \ket{\tilde{\underline{s}}_{b}}=
  \ket*{s_0,s_{\frac{1}{2}},\ldots,s_{\frac{L-1}{2}},b,s_{\frac{L+1}{2}},\ldots,s_{L-\frac{1}{2}}},
\end{equation}
and $\ket{\tilde{\underline{s}}_b^{t}}$ is its time-evolved counterpart,
\begin{equation}
  \ket{\tilde{\underline{s}}_{b}^t}=
  \mathcal{U}(t)\ket{\tilde{\underline{s}}_{b}}.
\end{equation}
Using this convention, we can express the vectorised version of the time-evolved operator $\mathcal{O}_{\frac{L}{2}}(t)$ in the computational basis as
\begin{equation}
  \ket*{\mathcal{O}_{\frac{L}{2}}(t)}=\sum_{\underline{s}}\sum_{b}
  \mel*{b}{\mathcal{O}}{s_{\frac{L}{2}}}
  \ket*{\tilde{\underline{s}}_{b}^t}\otimesR \ket{\underline{s}^t}.
\end{equation}
To be able to express purity we also introduce $\underline{s}_{S}$ to denote the restriction of configuration $\underline{s}$ to the subsystem $S$. In particular, for us the relevant subsystems are (cf.\ main text) $\bar{A}=\{0,\frac{1}{2},\ldots,\frac{L-1}{2}+\ell\}$, and $A=\{\frac{L}{2}+\ell,\ldots,L-\frac{1}{2}\}$, so that we have
\begin{equation}
  \ket{\underline{s}}
  =\ket{\underline{s}_{\bar{A}}\underline{s}_{A}}
  =\ket{\underline{s}_{\bar{A}}}\otimes \ket{\underline{s}_{A}},\qquad
  \underline{s}_{\bar{A}}=\{s_0,\ldots,s_{\frac{L-1}{2}+\ell}\},\quad
  \underline{s}_{A}=\{s_{\frac{L}{2}+\ell},\ldots,s_{L-\frac{1}{2}}\}.
\end{equation}
This allows us to express the local purity for a given realization as,
\begin{equation}
  \mkern-10mu
  \begin{aligned}
    \mathcal{P}_{\mathcal{O}}(t)=q^{-4L}
    \smashoperator[l]{\sum_{\substack{\underline{a},\underline{b},\underline{c},\underline{d},\\
    \underline{e},\underline{f},\underline{g},\underline{h}}}}
    &\smashoperator[r]{\sum_{a^{\prime},b^{\prime},c^{\prime},d^{\prime}}}
    \mel*{a^{\prime}}{\mathcal{O}}{a_{\frac{L}{2}}}
    \mel*{b^{\prime}}{\mathcal{O}^{\dagger}}{b_{\frac{L}{2}}}
    \mel*{c^{\prime}}{\mathcal{O}}{c_{\frac{L}{2}}}
    \mel*{d^{\prime}}{\mathcal{O}^{\dagger}}{d_{\frac{L}{2}}} \\[-1em]
    &\times\Big(
    \bra*{\underline{f}_{\bar{A}} \underline{h}_{A}}\otimesR
    \bra*{\underline{e}_{\bar{A}} \underline{e}_{A}}\otimesR
    \bra*{\underline{e}_{\bar{A}} \underline{g}_{A}}\otimesR
    \bra*{\underline{f}_{\bar{A}} \underline{f}_{A}}\otimesR
    \bra*{\underline{h}_{\bar{A}} \underline{f}_{A}}\otimesR
    \bra*{\underline{g}_{\bar{A}} \underline{g}_{A}}\otimesR
    \bra*{\underline{g}_{\bar{A}} \underline{e}_{A}}\otimesR
    \bra*{\underline{h}_{\bar{A}} \underline{h}_{A}}
    \Big)\\
    &\times\Big(
    \ket*{\tilde{\underline{a}}^{t}_{a^\prime \bar{A}}
    \tilde{\underline{a}}^{t}_{a^\prime A}}\otimesR
    \ket*{\underline{a}^{t}_{\bar{A}}\underline{a}^{t}_{A}}\otimesR
    \ket*{\tilde{\underline{b}}^{t}_{b^\prime \bar{A}}
    \tilde{\underline{b}}^{t}_{b^\prime A}}\otimesR
    \ket*{\underline{b}^{t}_{\bar{A}}\underline{b}^{t}_{A}}\otimesR
    \ket*{\tilde{\underline{c}}^{t}_{c^\prime \bar{A}}
    \tilde{\underline{c}}^{t}_{c^\prime A}}\otimesR
    \ket*{\underline{c}^{t}_{\bar{A}}\underline{c}^{t}_{A}}\otimesR
    \ket*{\tilde{\underline{d}}^{t}_{d^\prime \bar{A}}
    \tilde{\underline{d}}^{t}_{d^\prime A}}\otimesR
    \ket*{\underline{d}^{t}_{\bar{A}}\underline{d}^{t}_{A}}
    \Big),
  \end{aligned}
  \mkern-10mu
\end{equation}
which can be straightforwardly reduced to
\begin{equation}
  \mkern-8mu
  \begin{aligned}
    \mathcal{P}_{\mathcal{O}}(t)=q^{-4L} 
    \smashoperator[l]{\sum_{\underline{a},\underline{b},\underline{c},\underline{d}}}
    \smashoperator[r]{\sum_{a^{\prime},b^{\prime},c^{\prime},d^{\prime}}}
    \mkern-8mu
    \mel*{a^{\prime}}{\mathcal{O}}{a_{\frac{L}{2}}}\mkern-8mu
    \mel*{b^{\prime}}{\mathcal{O}^{\dagger}}{b_{\frac{L}{2}}}\mkern-8mu
    \mel*{c^{\prime}}{\mathcal{O}}{c_{\frac{L}{2}}}\mkern-8mu
    \mel*{d^{\prime}}{\mathcal{O}^{\dagger}}{d_{\frac{L}{2}}}\mkern-8mu
    \braket*{\underline{b}^t_{\bar{A}}\underline{d}^t_{A}}
    {\tilde{\underline{a}}^t_{a^\prime}}\mkern-8mu
    \braket*{\underline{a}^t_{\bar{A}} \underline{c}^t_{A}}
    {\tilde{\underline{b}}^t_{b^\prime}}\mkern-8mu
    \braket*{\underline{d}^t_{\bar{A}} \underline{b}^t_{A}}
    {\tilde{\underline{c}}^t_{c^\prime}}\mkern-8mu
    \braket*{\underline{c}^t_{\bar{A}} \underline{a}^t_{A}}
    {\tilde{\underline{d}}^t_{d^\prime}}.
  \end{aligned}
  \mkern-8mu
\end{equation}
Using again the fact that time-evolution is a one-to-one mapping between computational-basis states, we can rewrite the above expression as a double sum over computational basis states,
\begin{equation}\label{eq:GenPurCB}
  \mathcal{P}_{\mathcal{O}}(t)=q^{-4L} 
  \sum_{\underline{x},\underline{y}}
  f_{\mathcal{O}}(\underline{x}^{-t},(\underline{x}_{\bar{A}}\underline{y}_{A})^{-t},
  \underline{y}^{-t},(\underline{y}_{\bar{A}}\underline{x}_{A})^{-t}),
\end{equation}
where the superscript $-t$ means \emph{backwards} time-evolution, and the function $f_{\mathcal{O}}$ contains the operator dependent part, as well as the appropriate orthogonality condition for the four backwards-evolved configurations,
\begin{equation}\label{eq:GenPurCBf}
  f_{\mathcal{O}}(\underline{\alpha},\underline{\beta},\underline{\gamma},\underline{\delta})
  \mkern-3mu=\mkern-6mu
  \smashoperator{\sum_{a^{\prime},b^{\prime},c^{\prime},d^{\prime}\in\mathbb{Z}_{q}}}
  \mkern-4mu
  \mel*{\alpha_{\frac{L}{2}}}{\mathcal{O}}{a^{\prime}}\!
  \mel*{\beta_{\frac{L}{2}}}{\mathcal{O}^{\ast}}{b^{\prime}}\!
  \mel*{\gamma_{\frac{L}{2}}}{\mathcal{O}}{c^{\prime}}\!
  \mel*{\delta_{\frac{L}{2}}}{\mathcal{O}^{\ast}}{d^{\prime}}
  \braket*{\tilde{\underline{\alpha}}^t_{a^{\prime}\bar{A}}}{\tilde{\underline{\beta}}_{b^{\prime}\bar{A}}^t}\!
  \braket*{\tilde{\underline{\gamma}}^t_{c^{\prime} A}}{\tilde{\underline{\beta}}_{b^{\prime}A}^t}\!
  \braket*{\tilde{\underline{\gamma}}^t_{c^{\prime} \bar{A}}}{\tilde{\underline{\delta}}_{d^{\prime},\bar{A}}^t}\!
  \braket*{\tilde{\underline{\alpha}}^t_{a^{\prime} A}}{\tilde{\underline{\delta}}_{d^{\prime}A}^t}.
\end{equation}
For a given pair of configurations $(\underline{x},\underline{y})$, the factor $f_{\mathcal{O}}(\underline{x}^{-t},(\underline{x}_{\bar{A}}\underline{y}_{A})^{-t},\underline{y}^{-t},(\underline{y}_{\bar{A}}\underline{x}_{A})^{-t})$ is easy to evaluate, as it involves at most $q^4$ different terms whose complexity scales polynomially with $t$.

We can therefore estimate the averaged purity by sampling $\mathcal{N}$ different pairs of configurations $(\underline{x},\underline{y})$, and different circuit realizations $\mathcal{U}(t)$. For simplicity, and to avoid the sign problem, we focus on the case when $\mathcal{O}=\tau$ is the  shift operator $\tau\ket{s}=\ket{s+1 \pmod{q}}$. In this case, $f_{\mathcal{O}}$ can only take values $0$ or $1$, and for $q>2$ we observe $f_{\mathcal{O}}=1$ for an exponentially small number of samples. This gives linearly growing LOE, and limits the smallest possible accessible purity: in our case we used up to $10^{10}$ samples to resolve purities bigger than $10^{-9}$.

\subsection{Algorithm for diagonal operators}
For a diagonal operator $\mathcal{O}$ the discussion proceeds analogously, but the function in Eq.~\eqref{eq:GenPurCBf} simplifies, as the sum simplifies (operator forces only diagonal elements to survive), and in this case the orthogonality conditions are automatically fulfilled so that we have
\begin{equation}\label{eq:PurityGenDiagonalNum}
  \mathcal{P}_{\mathcal{O}}(t)=q^{-4L}\sum_{\underline{x},\underline{y}} 
  f^{\mathrm{d}}_{\mathcal{O}}
  (\underline{x}^{-t},(\underline{x}_{\bar{A}}\underline{y}_{A})^{-t}, \underline{y}^{-t},(\underline{y}_{\bar{A}}\underline{x}_{A})^{-t}),
\end{equation}
with
\begin{equation}
  f^{\mathrm{d}}_{\mathcal{O}}
  (\underline{\alpha},\underline{\beta},\underline{\gamma},\underline{\delta})=
  \mel*{\alpha_{\frac{L}{2}}}{\mathcal{O}}{\alpha_{\frac{L}{2}}}
  \mel*{\beta_{\frac{L}{2}}}{\mathcal{O}^{\ast}}{\beta_{\frac{L}{2}}}
  \mel*{\gamma_{\frac{L}{2}}}{\mathcal{O}}{\gamma_{\frac{L}{2}}}
  \mel*{\delta_{\frac{L}{2}}}{\mathcal{O}^{\ast}}{\delta_{\frac{L}{2}}}.
\end{equation}
When we randomly sample the sum in~\eqref{eq:PurityGenDiagonalNum}, most of the samples give non-zero contributions, but they might come with a negative sign, leading to the sign problem. In our numerics, we used up to $10^{10}$ samples, which allowed us to resolve purities bigger than $10^{-5}$.

\subsection{Algorithm for dynamics with phases}
If the dynamics contains also phases (cf.\ Sec.~\ref{sec:SuppPhases}), we proceed analogously. The main difference with respect to the original case is that in Eq.~\eqref{eq:GenPurCBf} every term in the sum needs to be multiplied by $\mathrm{e}^{\mathrm{i} \Phi}$ where the phase $\Phi$ is
\begin{equation}
  \Phi=\phi_t(\underline{\alpha})-\phi_t(\tilde{\underline{\alpha}}_{a^{\prime}})
  +\phi_t(\underline{\beta})-\phi_t(\tilde{\underline{\beta}}_{b^{\prime}})
  +\phi_t(\underline{\gamma})-\phi_t(\tilde{\underline{\gamma}}_{c^{\prime}})
  +\phi_t(\underline{\delta})-\phi_t(\tilde{\underline{\delta}}_{d^{\prime}}),
\end{equation}
an $\phi_t(\underline{s})$ is the shorthand for the sum of all the phases obtained by evolving the state $\ket{\underline{s}}$ up to time $t$,
\begin{equation}
  \mathcal{U}(t)\ket{\underline{s}}=\mathrm{e}^{\mathrm{i} \phi_t(\underline{s})}.
\end{equation}
From here it immediately follows that for the diagonal contributions in Eq.~\eqref{eq:GenPurCBf} there are no non-trivial contributions coming from phases, as we already observed in Sec.~\ref{sec:SuppPhases}.

With this generalisation we can again use random sampling of configurations to get a numerical estimate of the averaged LOE purity. As we show in the main text, the results are very similar, with the main difference occurring at $q=2$. Indeed, for $q=2$, two-site permutations with phases are no longer Clifford, and therefore LOE can (and does) grow linearly. For $q>2$ the differences in LOE are minimal, as for the majority of non-zero samples the phases cancel out.

\end{document}